\documentclass[twocolumn,superscriptaddress,amsmath,amssymb,showpacs,floatfix,preprintnumbers]{revtex4-1}

\usepackage{amsmath}
\usepackage{amssymb}
\usepackage{graphicx}
\usepackage{color}
\usepackage[colorlinks,citecolor=blue,linkcolor=blue]{hyperref}
\usepackage{epsfig}
\usepackage{amsmath}
\usepackage{amssymb}
\usepackage{epstopdf}
\usepackage{sidecap}
\usepackage{booktabs}
\usepackage{rotating}
\usepackage[bottom]{footmisc}
\usepackage[T1]{fontenc}

\sidecaptionvpos{figure}{c}

\DeclareGraphicsExtensions{.png .jpg .pdf}

\newcommand{\rmi}{\mathrm{i}}
\newcommand{\e}{\mathrm{e}}

\newcommand{\s}{\sigma}
\newcommand{\mc}{\mathcal}

\newcommand{\mr}{\mathrm}

\newcommand{\ibmstr}{\texttt{ibm\_strasbourg}}
\newcommand{\ibmfez}{\texttt{ibm\_fez}}

\begin{document}
\title{Quantum simulation of the Hubbard model on a graphene hexagon: 
	Strengths of IQPE and noise constraints}

\author{Mohammad Mirzakhani}\email{mohamad.mirzakhani@gmail.com}
\affiliation{Department of Physics, Yonsei University, Seoul 03722, Korea}

\author{Kyungsun Moon}\email{kmoon@yonsei.ac.kr}
\affiliation{Department of Physics, Yonsei University, Seoul 03722, Korea}
\affiliation{Institute of Quantum Information Technology, Yonsei University, Seoul 03722, Korea}

\begin{abstract}
Quantum computing offers transformative potential for simulating real-world materials, providing 
a powerful platform to investigate complex quantum systems across quantum chemistry and 
condensed matter physics.
In this work, we leverage this capability to simulate the Hubbard model on a six-site graphene
hexagon using Qiskit, employing the Iterative Quantum Phase Estimation (IQPE) and adiabatic 
evolution algorithms to determine its ground-state properties. 
Our results show that a single Slater determinant is sufficient to initialize IQPE and accurately recover ground-state energies (GSEs) in small-scale Hubbard systems. 
In noiseless simulations, IQPE converges within a few iterations to exact GSEs, while adiabatic simulations yield charge and spin densities and correlation functions in excellent agreement with exact diagonalization.
However, deploying IQPE and adiabatic evolution on today's noisy quantum hardware remains highly
challenging. 
To investigate these limitations in IQPE, we use the Qiskit Aer simulator with a custom noise model 
tailored to the characteristics of IBM's real hardware. 
This model includes realistic depolarizing gate errors, thermal relaxation, and readout noise,
allowing us to explore how these factors degrade simulation accuracy. 
Further, we implement the IQPE algorithm on IBM's \ibmstr\ and \ibmfez\ devices for a reduced three-site Hubbard model, enabling direct comparison between simulated and real hardware noise. 
While \ibmfez\ runs closely match exact results, discrepancies highlight the gap between modeled and physical noise. 
This study demonstrates both the IQPE's potential and current limitations for simulating strongly correlated systems under realistic conditions.
\end{abstract}

\maketitle


\section{Introduction} \label{s_intro}

The intersection of quantum computing and materials science has emerged as a promising avenue for
groundbreaking discoveries \cite{Georgescu2014,Bauer2016,Bauer2020}. 
By leveraging the power of quantum algorithms, researchers aim to simulate complex quantum systems, 
such as those found in novel materials, with unprecedented accuracy and efficiency. 
One such system of significant interest is the Fermi-Hubbard model  \cite{Hubbard1963}, 
a fundamental model in condensed matter physics that describes the behavior of 
interacting electrons on a lattice.

The Fermi-Hubbard model has long served as a fundamental framework for understanding 
the complex behavior of strongly correlated electron systems
\cite{Essler2005,Scalapino2007,Scalapino2012,LeBlanc2015,Qin2022}.
It captures key phenomena such as magnetism, superconductivity, and Mott insulator transitions
\cite{Mott1949}, making it essential for exploring the physics of high-temperature superconductors, 
quantum spin liquids \cite{Lee2008}, and beyond.
Despite its simplicity, solving the Hubbard model exactly for large systems remains a formidable 
challenge due to the exponential growth of the Hilbert space with system size. 
As of today, the largest Fermi-Hubbard system that has been solved exactly comprises 17 
electrons on a 22-site lattice \cite{Yamada2005}.
Conventional computational methods, such as density matrix renormalization group and quantum 
Monte Carlo, often face limitations, especially when addressing systems with strong correlations 
and intricate geometries \cite{Schollwock2005,Stoudenmire2011}.

Quantum computing offers a new way to tackle the Fermi-Hubbard model, as it can potentially represent 
the full Hilbert space more efficiently, opening the door to studying larger, more realistic models 
that are classically intractable.
The Fermi-Hubbard model has been extensively studied as a potential target for quantum simulation
\cite{Somma2002,Wecker2015s,Wecker2015p,Jiang2018,Reiner2019,Cade2020,Cai2020,Demers2020,Martin2022,
	Gard2022,Stanisic2022,Ayral2023}.
Its regular lattice structure and well-defined Hamiltonian, while presenting significant challenges 
for classical methods, make it an attractive benchmark for quantum algorithms.

Recent quantum simulations, particularly in quantum chemistry and materials science, 
have widely employed VQE (Variational Quantum Eigensolver) \cite{Cerezo2021,Alexeev2024} and 
QAOA (Quantum Approximate Optimization Algorithm) \cite{Farhi2014} 
for quantum many-body systems like the Hubbard model
\cite{Stanisic2022,Martin2022,Tilly2022,Blekos2024}. 
Such quantum-classical hybrid algorithms have been developed to overcome certain limitations.
They involve training parameterized quantum circuits to minimize the energy of the system.
For instance, VQE is an algorithm that leverages the power of quantum
computers for optimization tasks and is particularly useful for finding ground-state energy (GSE). 
However, its practical application is limited by the high classical computational overhead, 
especially when advanced error mitigation is implemented.

On the other hand, recent advancements in quantum hardware open up significant opportunities.
We may soon witness practical applications of quantum computing and the testing of fault-tolerant 
algorithms that were previously unattainable.
While encouraging the development of fault-tolerant quantum algorithms, these advancements 
also encourage a critical evaluation of near-term quantum approaches, exploring their 
potential utility across various scenarios.
Among these, Quantum Phase Estimation (QPE) \cite{Nielsen2010} stands out as a cornerstone
algorithm for realizing many quantum computing applications. 
QPE is crucial for tasks such as calculating molecular energies in quantum chemistry, solving
eigenvalue problems, and advancing materials science--all of which benefit from the precision 
and scalability offered by fault-tolerant quantum computing.

While less directly applied to the Hubbard model, QPE is also particularly well-suited for
determining the ground-state properties of quantum systems
\cite{Wecker2015s,Kitaev1995,Abrams1997,Abrams1999,Guzik2005,Whitfield2011}.
Combining the strengths of phase estimation and iterative algorithms, iterative QPE (IQPE) provides
a resource-efficient method for approximating GSEs 
\cite{Whitfield2011,Parker2000,Dobsicek2007,Mei2007}.
Abrams and Lloyd \cite{Abrams1997} were the pioneers in exploring time evolution governed by
the Hubbard Hamiltonian, albeit without presenting concrete quantum circuit implementations.
In Ref.~\cite{Abrams1999} they provide further details that the use of QPE can project trial 
states onto the corresponding energy eigenstates.
Technical details on simulating the Fermi-Hubbard model using phase estimation algorithm can 
be found in Refs.\ \cite{Wecker2015s,Whitfield2011}.

In this work, we focus on simulating the Hubbard model on a six-site graphene hexagon, a minimal
yet representative unit that encapsulates the material's correlated electron behavior. 
Leveraging Qiskit \cite{Qiskit}, an open-source quantum computing framework, we employ the IQPE 
algorithm to probe the system's ground-state properties with high precision in a noiseless environment. 
This approach, complemented by adiabatic time evolution, allows us to compute GSEs, charge and spin 
densities, and charge and spin correlations, providing a detailed understanding of 
graphene's electronic structure. 
To validate the power of this quantum method, we benchmark our results against exact
diagonalization, achieving excellent agreement and underscoring its capability 
to accurately model real materials.

Yet, the transition from idealized simulations to practical quantum hardware introduces significant
hurdles. 
IQPE, while powerful, demands deep circuits and precise control, making it particularly
sensitive to noise in noisy intermediate-scale quantum (NISQ) devices. 
For the system studied here, effectively a one-dimensional (1D) Hubbard chain with periodic boundary conditions (PBC), we demonstrate that the non-local Jordan-Wigner (JW) string arising from PBC can be simplified to a global phase factor that depends solely on the total fermion number in the fixed-occupancy sector.
This simplification significantly reduces circuit depth and noise, since 
implementing the original JW string requires multiple CNOT gates.

To investigate noise effects, we utilize the Qiskit Aer simulator \cite{Qis-Aer} with a 
noise model tailored to IBM's superconducting qubit platform \cite{IBM}, incorporating gate-specific 
depolarizing errors, thermal relaxation, and readout noise. 
For the noisy simulations emulating IBM QPUs, we reduce the system register to $N = 3$ atomic sites.
This qubit reduction keeps simulations computationally efficient while allowing a detailed examination of noise effects on IQPE performance.
This noisy simulation reveals both the robustness of our results and their degradation 
under realistic conditions. 

Further, we implement the IQPE algorithm on real IBM quantum hardware,
employing the \ibmstr\ and \ibmfez\ devices for a reduced system size of $N = 3$ atomic sites.
This allows for a direct comparison between simulated noise models and actual hardware performance,
revealing discrepancies between simulated and physical noise behavior.
Notably, runs on the \ibmfez\ device yield GSEs in excellent agreement with the exact results.
This work provides a step toward benchmarking quantum algorithms for strongly correlated systems and serves as a case study for evaluating IQPE performance on NISQ devices.

\section{Theory and model} \label{s_the_mod}
\subsection{Graphene fragments and the Hubbard Model} \label{ss_grap_hub}

The Hubbard model, a fundamental framework in condensed matter physics, captures the essential
physics of electrons in a lattice by incorporating electron-electron (e-e) interactions and hopping
between neighboring sites \cite{Essler2005,LeBlanc2015,Qin2022}. 
It is particularly effective in describing strongly correlated phenomena such as magnetism,
metal-insulator transitions, and superconductivity. 
In the context of graphene--a material with a hexagonal lattice where each carbon atom contributes
a $p_z$ orbital to the delocalized $\pi$-band--the Hubbard model provides a natural description 
of its low-energy electronic and magnetic properties \cite{Geim2007,Castro2009}. 
For finite graphene structures, often referred to as nanoflakes or quantum dots, adaptations of 
the tight-binding and Hubbard models are used to capture size and edge effects, which play a
crucial role in determining the system's electronic behavior
\cite{Rozhkov2011,Oteyza2022,Fernandez2007,Guclu2014,Yazyev2010}.
Here, we consider the smallest possible graphene fragment: a single hexagonal unit of the graphene 
honeycomb lattice, consisting of six carbon atoms positioned at its vertices, as illustrated in
Fig.~\ref{fig1}(a).
Notice that this compact structure can also be considered as a 1D Hubbard chain with PBC.

The Hamiltonian of the Hubbard model consists of two main terms that describe the competing 
effects of electron hopping ($H_0$) and on-site e-e repulsion ($H_\mr{U}$) \cite{Hubbard1963}:
\begin{equation}  \label{eq:Htot}
	\mc H = \mc{H}_0 + \mc{H}_\mr{U}.
\end{equation}
The first term, $H_0$, is the single-particle tight-binding Hamiltonian, representing 
the hopping of electrons between adjacent lattice sites. 
In the second quantization formalism, it can be written as
\begin{equation}  \label{eq:h0}
	\mc{H}_0 = \sum_{i,\s} \epsilon_{i \s} a^\dagger_{i \s} a_{i \s}
	-\gamma_0 \sum_{\langle i,j \rangle, \s} \left(a^\dagger_{i \s} a_{j \s} 
	+ \text{h.c.}\right).
\end{equation}
Here, $a^\dagger_{i \s}$ and $a_{j \s}$ are the electron creation and annihilation operators, 
respectively, which add or remove an electron with spin $\s$ at sites $i$ and $j$. 
$\epsilon_{i \s}$ and $\gamma_0$ are the on-site and hopping energies, respectively. 
Here, we set $\epsilon_{i \s} = 0$ and $\gamma_0 = 1$.
The sum $\langle i,j \rangle$ restricts this interaction to nearest-neighbor sites.

The second term in Hamiltonian \eqref{eq:Htot} is the Hubbard term that introduces 
e-e interactions through the repulsive on-site Coulomb interaction,
\begin{equation}  \label{eq:Hub}
	\mc{H}_\mr{U} = U_0 \sum_i n_{i \uparrow} n_{i \downarrow},
\end{equation}
where $n_{i \s} = a^\dagger_{i \s} a_{i \s}$ ($\s = \uparrow, \downarrow$) is the 
spin-resolved electron density at site $i$.
The parameter $U_0 > 0$ is the Hubbard parameter and denotes, in the short-range
regime, the on-site Coulomb repulsion energy for each pair of electrons with opposite 
spins on the same site $i$. 

\subsection{Quantum computing approach} \label{ss_qu_co_app}
QPE is a quantum algorithm that enables us to efficiently estimate the eigenvalues of a given
Hamiltonian \cite{Whitfield2011,Guzik2005,Wecker2015s}.
This capability is essential for finding the ground state and low-energy excited states of a system, 
which are critical for determining its electronic and magnetic properties.
IQPE is a simplified and resource-efficient variant of QPE \cite{Dobsicek2007,Parker2000,Whitfield2011}. 
Unlike standard QPE, IQPE estimates the phase iteratively, qubit by qubit, and does not require an
ancillary qubit register for each bit of precision.
This makes it more practical for near-term quantum devices with limited qubit resources.
In theory, the algorithm's accuracy is constrained only by the number of iterations. 
However, in practice, it is limited by experimental imperfections. 
The highest experimentally achievable accuracy can act as a benchmark for evaluating the performance
and identifying limitations of IQPE algorithm implementations on current NISQ devices.
We take the following approach to simulate the Hubbard model on a quantum hardware.

\subsubsection{Preparing initial state} \label{sss_init_stat}
The success of the IQPE algorithm [Fig.~\ref{fig2} ]largely depends on the choice of the initial state, 
as it must have a significant overlap with the target eigenstate of the Hamiltonian 
being solved, here, the ground state of the Hubbard model.
However, achieving a perfect overlap with the ground state is not essential; 
if the overlap is sufficiently large, we can apply phase estimation algorithm to 
the state and still have a reasonably large chance of projecting onto the ground state.
%

\textit{Slater determinant (SD)}:
For fermionic systems like the Hubbard model, a natural choice for an initial state is an SD.
An SD can be regarded as an eigenstate of a quadratic Hamiltonian, or simply as 
a wavefunction that describes a set of non-interacting fermions that can be represented as a
product of single-particle states.
The standard method for preparing SDs was introduced in Ref.~\cite{Ortiz2001} and later optimized 
in Refs.~\cite{Wecker2015s,Kivlichan2018} using Givens rotations--elementary operations that perform
rotations in the plane defined by two coordinate axes.
Jiang \textit{et al.} in Ref.~\cite{Jiang2018} improved the existing algorithm to prepare an arbitrary 
SD by exploiting a unitary symmetry.

\textit{Annealed initial state}:
Although it is simple to prepare on a quantum computer, an SD might not always have a high overlap 
with the true correlated ground state of a strongly interacting system.
If the overlap with the target ground state is low, an \textit{annealed} state can be used as an
approximation of the ground state of the full Hubbard Hamiltonian.
This state generally has a much higher overlap with the actual ground state than a simple SD.
To prepare an annealed state as an initial state, one can use an adiabatic state preparation technique
\cite{Guzik2005,Farhi2001}. 
This process involves gradually evolving from the ground state of a known Hamiltonian, such as a 
SD from Hartree-Fock, to the ground state of the target Hubbard Hamiltonian.
To achieve this, we introduce a smoothly varying adiabatic path Hamiltonian:
\begin{equation}
	\mc H_\mr{ad}(\eta) = (1 - \eta(t)) \mc H_0 + \eta(t) \mc H, 
\end{equation}
where $0 \leq \eta(t) \leq 1$.
A common choice is $\eta(t)=t/T$, where $T$ is the total evolution time, though alternative schedules 
(e.g., nonlinear variations) can optimize the process.
Adiabatic state preparation is efficient when $T$ satisfies the following condition:
\begin{equation} \label{eq:T_evol}
	T \gg [\Delta_\mr{min} (\eta)] ^{-2},
\end{equation}
where $\Delta_\mr{min} (\eta)$ is the minimum energy gap between the ground state and the first excited 
state of $\mc H_\mr{ad}(\eta)$ during evolution.
For a more comprehensive discussion on state preparation techniques, including the adiabatic 
approach, see Refs.~\cite{Wecker2015s,Whitfield2011} and references therein.

Here, however, due to the small size of our system, we utilize SDs as the initial state, prepared
using the algorithm in Ref.~\cite{Jiang2018} and implemented in Qiskit Nature \cite{QisNat}.
Our numerical results demonstrate the effectiveness of these initial states in achieving reasonably
accurate results with the IQPE algorithm.

\subsubsection{Unitary time evolution of the Hubbard model} \label{sss_tim_evo}
The core of IQPE algorithm involves simulating the time evolution of the system under the Hubbard 
Hamiltonian \eqref{eq:Htot}, represented by the unitary operator 
$ \mc {U_H} (t) = \exp(- \rmi \mc{H} t) $.
This unitary evolution can be approximated using techniques like Trotterization 
\cite{Hatano2005}, which breaks down the exponential of the Hamiltonian into a sequence of gates.
This section outlines the Trotter decomposition for non-commuting Hamiltonian terms and provide
explicit quantum circuit constructions for the relevant exponentials.

\textit{Trotter-Suzuki decomposition}:
Trotterization, often referred to as Trotter-Suzuki decomposition, is a key technique in quantum
computing used to approximate the time evolution of quantum systems by decomposing the total
evolution into a product of evolutions for individual Hamiltonian terms.
Suppose the Hamiltonian $ \mc H $ can be decomposed as a sum of terms 
$ \mc H = \mc H_1 + \mc H_2 + \dots + \mc H_k $, where each term $ \mc H_i $ represents a
distinct part of the system, such as interactions or kinetic energy components.
Because these terms do not generally commute, the first-order Trotter-Suzuki decomposition
approximates the evolution operator as:
\begin{equation}
	e^{-\rmi \mc H t} \approx \left( e^{-\rmi \mc H_1 \Delta t} e^{-\rmi \mc H_2 \Delta t} 
	\dots e^{-\rmi \mc H_k \Delta t} \right)^{N_\mr{trot}},
\end{equation}
where $ \Delta t = t /N_\mr{trot} $ and $N_\mr{trot}$ is the number of Trotter steps~\cite{Hatano2005}.
As $N_\mr{trot}$ increases, the approximation becomes more accurate.
By applying the Trotterized sequence of these smaller exponentiated terms, we can approximate the
evolution of our system on a quantum computer.
\begin{figure}
	\centering
	\includegraphics[width = 8.0 cm]{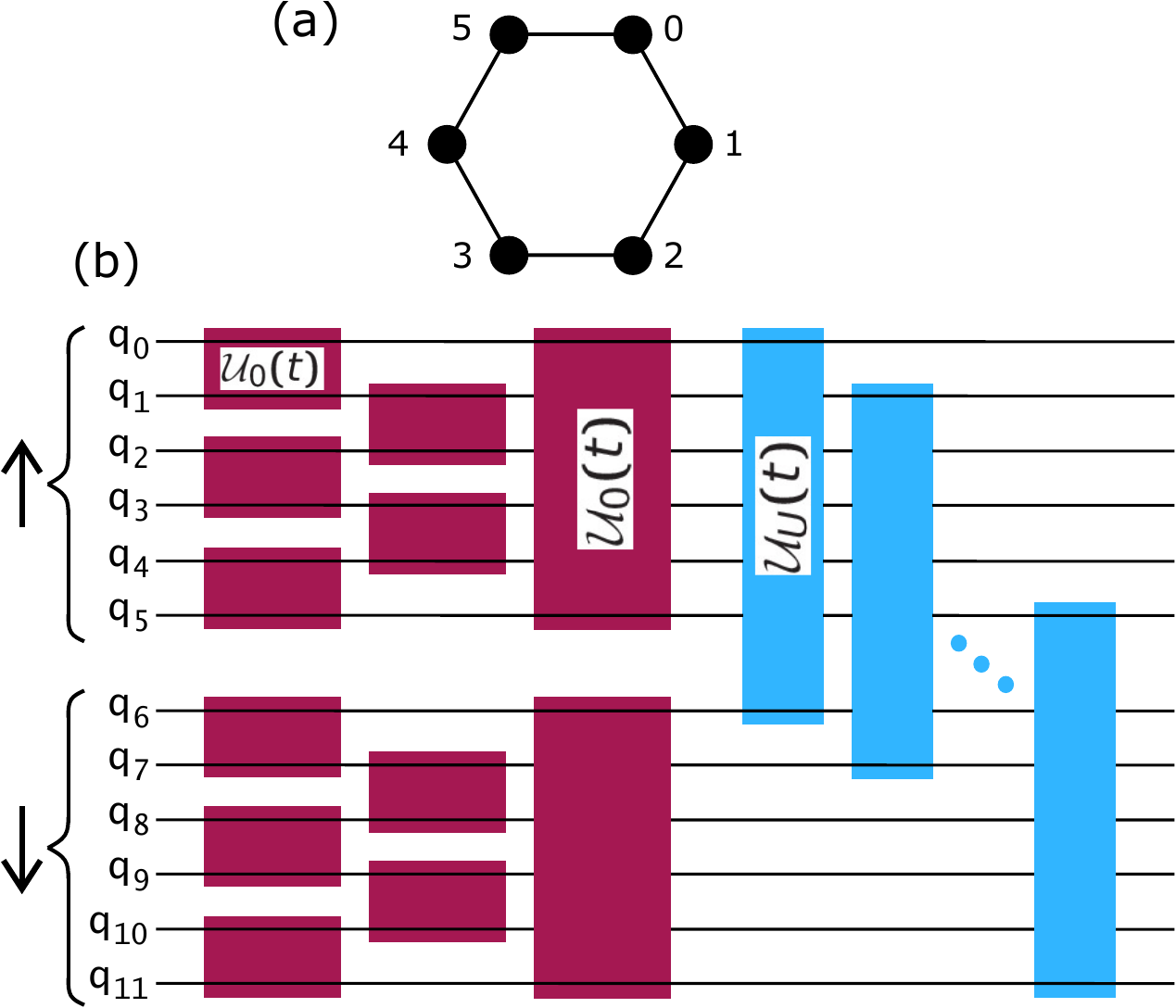}
	\caption{(a) Graphene hexagon with six atomic sites labeled from 0 to 5.
		(b) A schematic quantum circuit for the time evolution under the Hubbard model
		Hamiltonian \eqref{eq:Htot}.
		The first $N = 6$ qubits correspond to spin-up states, while the remaining $ N = 6$ qubits 
		represent spin-down states.
		The red boxes, $\mc U_0(t)$, implement the time evolution under the hopping terms between 
		two adjacent atomic sites. 
		The blue boxes, $\mc U_U(t)$, represent the time evolution under the on-site Coulomb 
		interaction	between qubits of opposite spins, i.e., $q_i$ and $q_{N+i}$. 
		Details of the unitary operators $\mc U_0(t)$ and $\mc U_U(t)$ are provided in the Appendix.
	}
	\label{fig1}
\end{figure}
\textit{Quantum circuits}:
In order to simulate a fermionic Hamiltonian on a quantum computer, one needs to use the 
well-known JW transformation, which maps each fermionic mode to a single
qubit arranged along a 1D line \cite{Ortiz2001,Jordan1928}.
In the JW transformation (see Appendix), each fermionic mode (which can be either occupied
or unoccupied) is mapped to a spin-$1/2$ system (a qubit), where the presence or
absence of a fermion corresponds to the qubit being in the $ | 1 \rangle $ or 
$ | 0 \rangle $ state.
The transformation uses Pauli operators ($X, Y, Z, I$) to capture the creation and annihilation of 
fermions, with careful handling of the anti-commutation relationships. 
Specifically, a string of Pauli $Z$ operators is introduced to maintain the correct
anti-commutation properties, which is essential for ensuring that fermionic operators act
properly in the quantum simulation.

In the case of Hubbard model, we need to implement unitary evolution under two types of terms:
hopping terms $ h_0 = -\gamma_0 (a^\dagger_{i \s} a_{j \s} + a^\dagger_{j \s} a_{i \s}) $ 
and on-site Coulomb interactions $ h_U = U_0 n_{i \s} n_{i \s'} $.
To establish the necessary notation, we present the explicit Pauli representation of the 
Hubbard model and outline the corresponding quantum circuits (see Appendix) used to solve 
the Hubbard model in this paper.

Using the JW transformation [Eq.~\eqref{eq:JW}] for the hopping terms $h_0$ in \eqref{eq:h0}, 
we can derive the following expression for the case $j > i$: 
\begin{equation} \label{eq:hop}
	(a^\dagger_{i} a_{j} + a^\dagger_{j} a_{i}) \mapsto 
	\frac{1}{2}(X_i X_j + Y_i Y_j) Z_\mr{JW},
\end{equation}
where $Z_\mr{JW} = \otimes_{k=i+1}^{k=j-1} Z_{k}$.
In the Hamiltonian $\mc H_0$ \eqref{eq:h0}, certain terms may act on more than two qubits 
through a $Z_\mr{JW}$ string (JW string) depending on the arrangement of the lattice sites. 
The corresponding quantum circuit for the time evolution under the hopping terms $h_0$, i.e.,
 $ \mc U_0(t) = \exp(-\rmi h_0 t) $, is provided in the Appendix.
Using a similar approach, we can express the number and Coulomb-interaction operators 
in terms of Pauli matrices, respectively: 
\begin{align}
	n_{i} &= a^\dagger_{i} a_{i} \mapsto \frac{1}{2}(I_i - Z_i), \label{eq:num} \\ 
	n_{i} n_{j} &\mapsto \frac{1}{4}(I_i I_j - I_i Z_j - Z_i I_j + Z_i Z_j). \label{eq:hub} 
\end{align}
For the circuits that implement the time evolution of these operators, refer to the Appendix.

For graphene fragments with $N$ atoms, the numbering of atomic sites shown in
Fig.~\ref{fig1}(a) enables us to map the lattice onto a qubit line.
We arrange the $2N$ qubits such that the first $N$ qubits, indexed by the lattice-site positions,
correspond to spin-up states and the remaining $N$ qubits correspond to spin-down states, so that
$ a^\dagger_{i \uparrow} = a^\dagger_{i} $ and $ a^\dagger_{i \downarrow} = a^\dagger_{i+N} $.
This ordering simplifies the JW strings required, as the hopping terms conserve spin.
We will return to this in more detail below.

With this pattern,  we demonstrate in Fig.~\ref{fig1}(b) the corresponding circuit for a graphene
hexagon with $N = 6$ that implements time evolution for a time step $t$ under the Hubbard 
Hamiltonian in Eq.~\ref{eq:Htot}.
In this circuit, $\mc U_0(t)$ (red boxes) implements the time evolution of the hopping terms between 
neighboring sites according to their ordered indices within each spin state.
The operation $ \mc U_U (t)$ ($\equiv \exp(-\rmi h_U t)$, blue boxes) corresponds 
to the on-site Coulomb interaction between spin-up and spin-down states, acting 
between pairs of qubits with opposite spins, specifically $q_i$ and $q_{i+N}$ 
where $i = 0, 1, \ldots, 5$.

Due to the PBC of our structure, a JW string appears between the first and last qubits within each spin sector (the rightmost red boxes).
This string increases the circuit depth and contributes to noise, as its implementation typically requires multiple CNOT gates.
However, we demonstrate that in a 1D Hubbard chain with PBC, the $Z_\mr{JW}$ string can be simplified and expressed in terms of a global phase factor depending on the fermionic occupation numbers $N_f$ in each spin sector as (see Appendix):
\begin{align} \label{eq:jw_per}
	(a_N^\dagger a_1 + \text{h.c.})
	& = \left[\left(\otimes_{k=1}^{N-1} Z_k\right) \otimes \sigma_N^\dagger \sigma_1
	+ \text{h.c.} \right] \notag \\
	& = (-1)^{N_f - 1} \Big[\sigma_1 \otimes \sigma_N^\dagger + \text{h.c.} \Big],
\end{align}
where $\sigma_j = \frac{1}{2}(X_j + \rmi Y_j)$ and $\sigma_j^\dagger = \frac{1}{2}(X_j - \rmi Y_j)$ are the spin lowering and raising operators at site $j$, respectively.

\begin{figure}
	\centering
	\includegraphics[width = 8.0 cm]{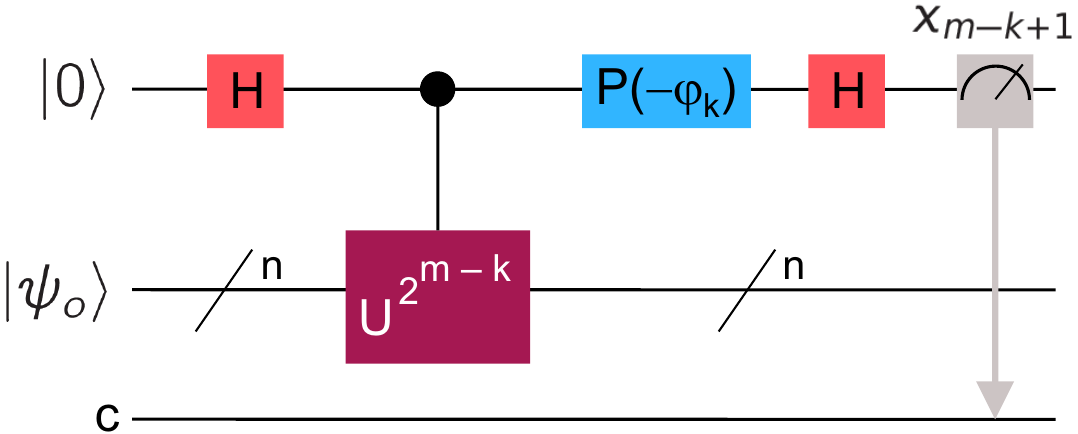}
	\caption{The IQPE algorithm for the $k$th iteration.
		In the circuit, $ |\psi_o \rangle $ is an eigenstate of $U$, and the angle 
		$\varphi_k$ of the phase gate depends on the previously measured bits defined 
		as $\varphi_k = 0.0 x_{m-k+2} x_{m-1} x_m$ binary, where $m$ is the number 
		of digits determining the accuracy of the phase $\phi_o$. 
		Note that $\varphi_k$ is zero in the first iteration of the algorithm.
		}
	\label{fig2}
\end{figure}

\subsubsection{IQPE} \label{sss:iqpe}
Here, we briefly describe the IQPE algorithm, which has been extensively detailed in many studies 
(see, e.g., Refs.\ \cite{Whitfield2011, Parker2000, Dobsicek2007}). 
IQPE is an adaptation of the QPE algorithm, widely used in quantum computing for tasks such as 
solving eigenvalue problems and simulating quantum systems \cite{Somma2002,Whitfield2011,Wecker2015s}.

As shown in Fig.~\ref{fig2}, the IQPE algorithm for solving an eigenvalue problem requires two 
inputs: (1) an initial state $ | \psi_o \rangle $ that is close to an eigenstate of the system's
Hamiltonian $\mc H$, and (2) a unitary operator $\mc U$ [$\equiv \exp(-\rmi \mc H t )$ with eigenvalue
$\exp(\rmi 2 \pi \phi_o)$], which represents the 
time evolution of the system.
The process begins by preparing the system in the eigenstate $ | \psi_o \rangle $ and
initializing an ancillary qubit in the state $ |+ \rangle = (|0 \rangle + |1 \rangle) / \sqrt{2} $
(the first Hadamard ($H$) gate in Fig.~\ref{fig2}).
A controlled-$\mc U^{2^{m-k}}$ operation is then applied, where $k\ = 1, 2, \ldots, m$ denotes the 
iteration index and $m$ represents the desired number of binary digits determining the precision 
of a phase parameter $\phi_o = 0.x_1 x_2 \ldots x_m$.
In the first iteration ($k = 1$), a controlled-$\mc U^{2^{m-1}}$ operation and $P(\varphi_1 = 0$) 
are applied.
The control qubit is then measured in the computational basis following the application of the 
second $H$ gate, yielding the first bit $x_m$ of the phase expansion.
Next, we update the phase estimate $\varphi_2 = 0.0 x_m$ for $P(\varphi_2)$,  
preparing for the second iteration ($k = 2$).
In the $k$th iteration, a controlled-$\mc U^{2^{m-k}}$ operation and $P(\varphi_k)$ are 
applied, where $\varphi_k = 0.0 x_{m-k+2} x_{m-1} x_m$, updated from previous iterations.
This measurement result is used to refine the phase estimate, determining $x_{m-k+1}$.
The algorithm concludes when the final bit $x_1$ is obtained in the $k=m$ iteration.
By iteratively repeating these steps and incrementing $k$ during each iteration, the 
algorithm successively refines the phase estimate, determining each bit with increasing accuracy.
Once the phase is estimated, the corresponding eigenvalue of the Hamiltonian can be computed as
$E_o = -2 \pi \phi_o / t$. 

\begin{figure}
	\centering
	\includegraphics[width = 8.5 cm]{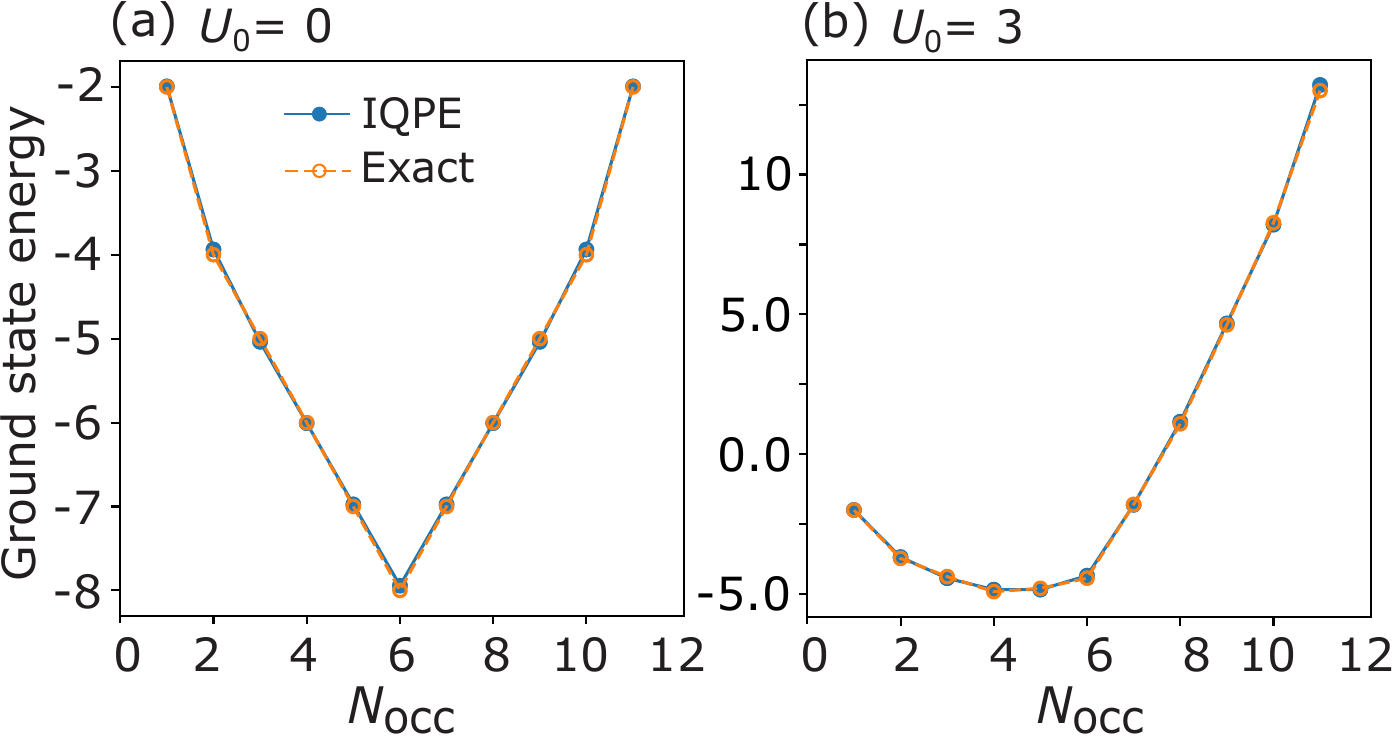}
	\caption{GSE as a function of occupation number $N_\mr{occ}$ for a Hubbard 
		model on a graphene hexagon lattice. 
		(a) Non-interacting case ($U_0=0$) and (b) and interacting case ($U_0=3$).
		Filled blue circles represent (noiseless) IQPE results, while open orange circles
		denote exact diagonalization.
		The close agreement between IQPE and exact results demonstrates the accuracy of the 
		quantum algorithm.
	}
	\label{fig3}
\end{figure}

\section{Numerical results} \label{s_con}

\subsection{Noiseless simulation} \label{ss_nois_less}
We begin by computing the GSE for all occupation numbers $1 \leq N_\mr{occ} \leq 11$, considering 
both the non-interacting case ($U_0=0$) and the presence of on-site interaction energy ($U_0 = 3$),
as shown in Figs.~\ref{fig3}(a) and \ref{fig3}(b), respectively.
We simulate our structure using the noiseless Qiskit Aer simulator and use exact diagonalization
results obtained with the QuSpin package \cite{QuSpin} as a benchmark for comparison.
In the case of the IQPE algorithm, we obtain results with a binary precision of $m=5$ digits.
To accurately recover $E_o$, $t$ must be chosen such that the phase $E_o t$ remains in the range
$[0, 2\pi)$ (or equivalently $\phi_o \in [0, 1)$) to avoid phase wrapping and ambiguity.
The exact value of $t$ depends on the energy scale of the system; often it is chosen such that even 
the maximum eigenvalue of interest satisfies $E_\mr{max} t < 2 \pi$.
However, if $t$ is too large, the phase might wrap around (i.e., become modulo $2\pi$), leading 
to ambiguities. 
Therefore, $t$ must be chosen based on an estimate of the energy spectrum of $\mc H$.
In practice, one might choose $t$ based on prior knowledge of the system's energy scale or by
performing preliminary runs to determine a suitable scaling. 
For the time evolution under $\mc {U_H} (t)$, we employ a Trotterization scheme with 
$N_\mr{trot} = 15$ steps as explained in Sec.~\ref{ss_qu_co_app}.

Figure \ref{fig3} demonstrate the GSE as a function of occupation number $N_\mr{occ}$ for a graphene
hexagonal lattice described with the Hubbard model. 
We observe a strong quantitative agreement between the exact diagonalization (orange open circles) 
and IQPE results (blue filled circles), demonstrating the accuracy and reliability of the IQPE method 
in capturing GSEs.
In the non-interacting case ($U_0=0$), Fig.~\ref{fig3}(a) shows a symmetric V-shaped dependence centered at half-filling ($N_\mathrm{occ}=6$),
reflecting the particle-hole symmetry of the kinetic term.
In this case, as seen, the half-filling configuration has the lowest GSE. 
In the interacting case [$U_0=3$, Fig.~\ref{fig3}(b)], GSE shows a significant deviation from the
non-interacting trend, with a flatter minimum and an upward energy shift due to the on-site 
Coulomb repulsion. 
The lowest GSE varies with the interaction strength $U_0$, occurring at different occupation numbers
depending on its magnitude. 
For instance, at $U_0=3$, the lowest GSE is found for $N_\mr{occ}=4$ as seen in Fig.~\ref{fig3}(b).
Despite the presence of interactions, IQPE still remains highly accurate, while using the SD as 
the initial state. 
These findings highlight the capability of IQPE in accurately determining eigenvalues,
demonstrating its potential for simulating strongly correlated systems using quantum computing.

\begin{figure}
	\centering
	\includegraphics[width = 8.5 cm]{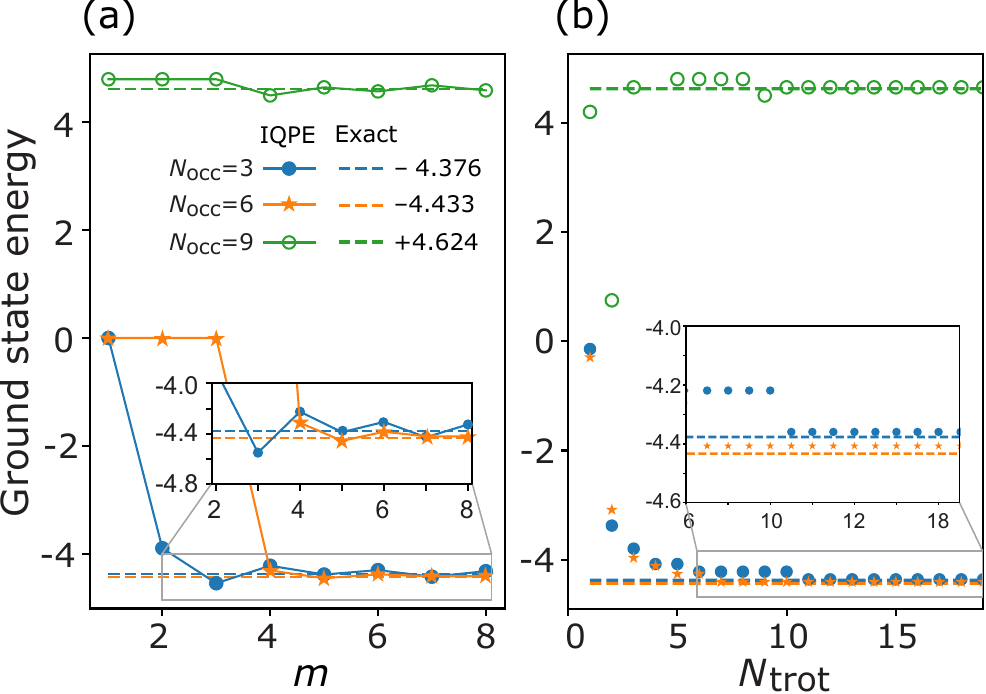}
	\caption{Convergence of the GSE as a function of (a) bit precision $m$ with $N_\mr{trot} = 15$ 
		and (b) Trotter step number $N_\mr{trot}$ with $m=5$ for three different occupation numbers 
		$N_\mr{occ} = 3,\ 6$ and $9$.  
		Results are shown for the interacting case with $U_0 = 3$.
	}
	\label{fig4}
\end{figure}

Below, we analyze the impact of bit precision $m$ and Trotter step number $N_\mr{trot}$ on 
the accuracy of IQPE results.
In Figs.~\ref{fig4}(a) and \ref{fig4}(b), we depict the convergence of the GSE as a function of binary
precision $m$ (with $N_\mr{trot} = 15$) and Trotterization steps $N_\mr{trot}$ (with $m=5$) for 
three different occupation numbers $N_\mr{occ} = 3,\ 6,$ and $9$, respectively.
The on-site potential is $U_0 = 3$.
As seen, only within a few iterations, IQPE algorithm is enable to converge to the exact value 
with the high accuracy.
For the structure studied here, selecting optimal values of $N_\mr{trot} = 15$ and $m = 5$ yields
highly accurate (noiseless) results, which are also feasible for execution on current quantum hardware.

\begin{figure}
	\centering
	\includegraphics[width = 8 cm]{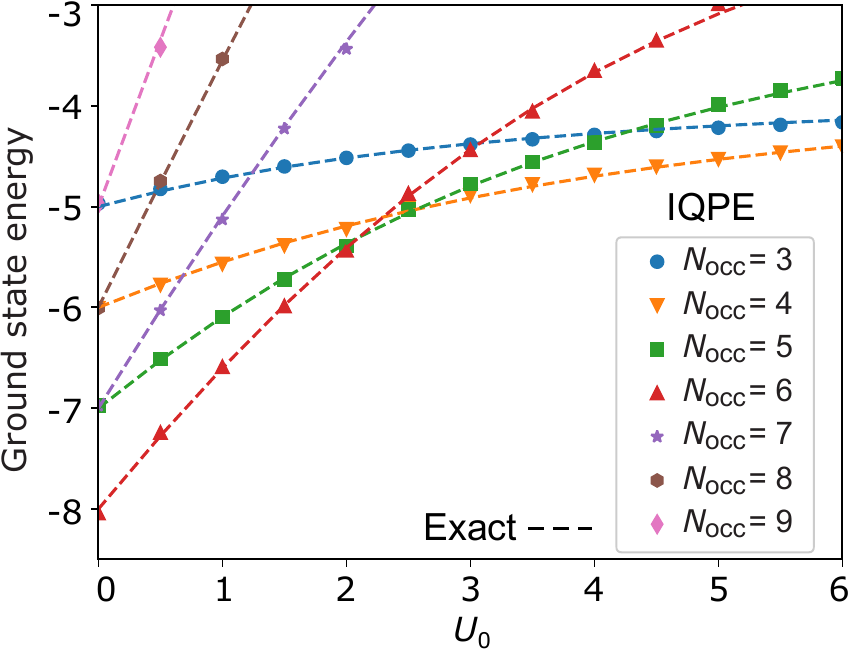}
	\caption{The GSE as a function of the on-site interaction strength $U_0$ for occupation 
		numbers  $3 \leq N_\mr{occ} \leq 9$.
		Symbols represent IQPE results, while the dashed curves correspond to exact 
		diagonalization results.
		The IQPE results are obtained using a binary precision of $m=5$ and Trotter step number of 
		$N_\mr{trot} = 15$.
	}
	\label{fig5}
\end{figure}

Next, we study the dependence of the GSE on the on-site potential, $U_0$, for a range of 
occupation numbers, $3 \leq N_\mr{occ} \leq 9$ , as presented in Fig.~\ref{fig5}.
Symbols represent the IQPE results, while dashed lines indicate the exact diagonalization outcomes.
The excellent agreement between IQPE and exact results demonstrates the accuracy and reliability of the 
IQPE algorithm in capturing the GSE of the Hubbard model across various interaction 
strengths and particle fillings.
As expected, the GSE generally increases with increasing on-site interaction strength. 
This behavior arises from the repulsive nature of the on-site interaction term, which penalizes
configurations with double occupancy of lattice sites.
The energies for different $N_\mr{occ}$ exhibit distinct behavior. 
For larger $N_\mr{occ}$, the energy increases more rapidly with $U_0$ compared to smaller $N_\mr{occ}$. 
This is expected since more electrons lead to stronger interaction effects.
As seen, for small $U_0$, energy differences are larger, but as $U_0$ increases, the trends become more parallel,
suggesting that strong interactions dominate over kinetic contributions.
For $U_0/ \gamma_0 \lessapprox 2 $, the half-filled configuration exhibits the lowest GSE.
However, beyond $U_0 \approx 2$,  the configurations with $N_\mr{occ} = 5$ (for a short range) and $N_\mr{occ} = 4$ 
become the lowest-energy states as $U_0$ increases.

Following, we analyze the behavior of charge and spin densities ($n_i,\ s_i$) along with the 
charge and spin correlations ($C^c_{i, j},\ C^s_{i, j}$), which are respectively defined as:
\begin{align}
	 n_i  &= n_{i \uparrow} + n_{i \downarrow}, \\
	S_i^z &= n_{i \uparrow} - n_{i \downarrow}, \\
	C^c_{i, j} &= \langle n_i n_j \rangle - \langle n_i \rangle \langle n_j \rangle, \\
	C^s_{i, j} &= \langle S_i^z S_j^z \rangle - \langle S_i^z \rangle \langle S_j^z \rangle.
\end{align}
For this purpose, we utilize the adiabatic evolution method introduced in Sec.~\ref{sss_init_stat}.
In this study, we consider a linear evolution path, given by $\eta (t) = t/T$.

\begin{figure}
	\centering
	\includegraphics[width = 8.5 cm]{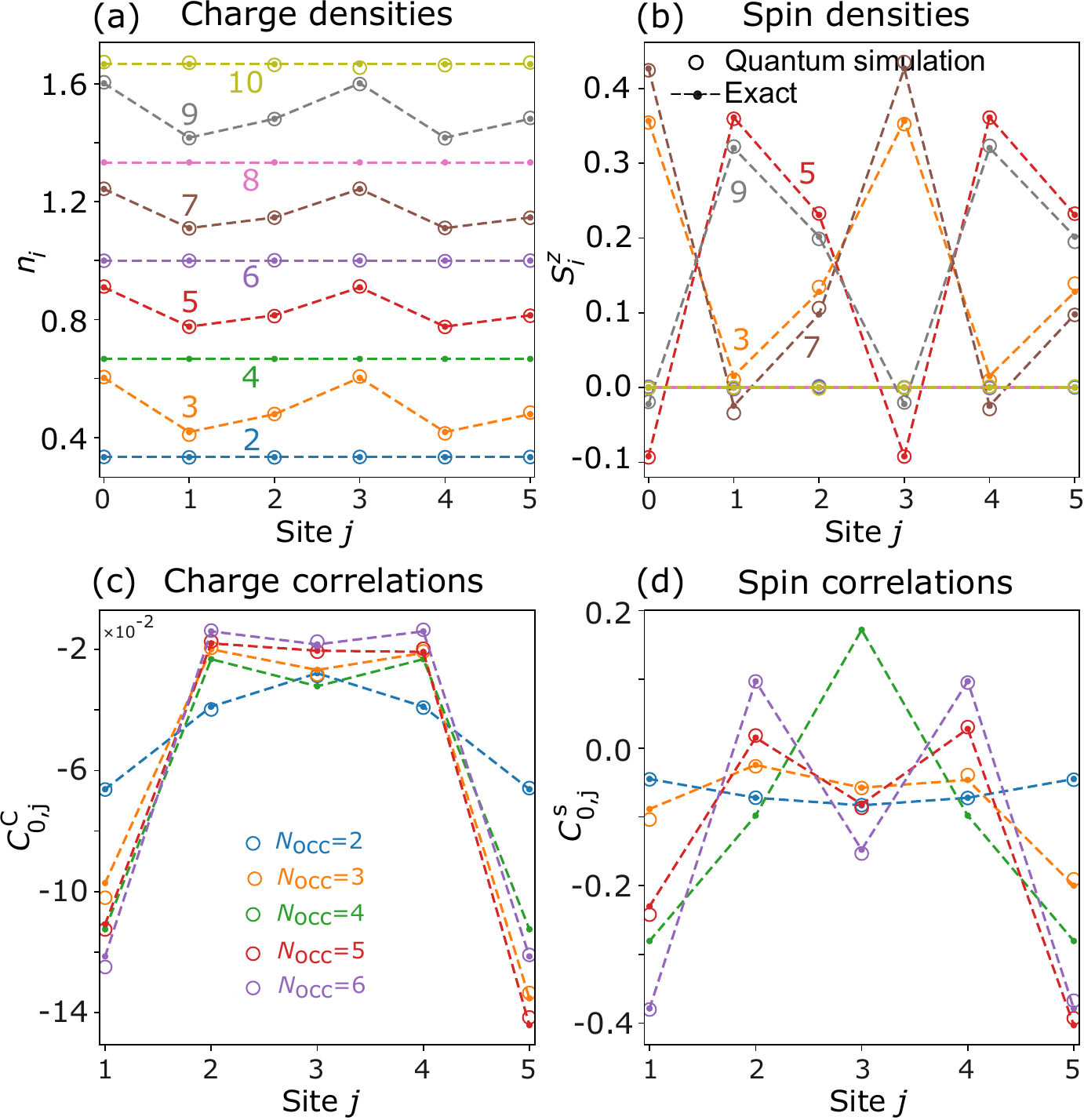}
	\caption{Comparison of (a) charge densities $n_i$, (b) spin densities $S_i^z$,
		(c) charge correlations $C_{0,j}^c$, and (d) spin correlations $C_{0,j}^s$ 
		obtained from adiabatic quantum simulation 
		(empty circles) and exact diagonalization (dashed lines). 
		The interaction strength is $U_0 = 3$.  
		Results are presented for different occupation numbers, as indicated in each panel.
		The excellent agreement between methods demonstrates the accuracy of quantum simulation 
		in capturing electronic correlations. 
	}
	\label{fig6}
\end{figure}

Figure \ref{fig6} shows the distribution of (a) charge densities and (b) spin densities across the 
six atomic sites for different occupation numbers ($2 \leq N_\mr{occ} \leq 10$) at $U_0=3$.
The dashed lines represent the exact diagonalization densities while the empty circles correspond to the
adiabatic-evolution results.
As seen, the quantum simulation results closely follow the exact diagonalization results for all
$N_\mr{occ}$. 
This suggests that the quantum annealing algorithm is performing well in capturing the ground state of 
the system, at least for this small system size.
Charge densities for all odd occupation numbers exhibit similar trends [Fig.~\ref{fig6}(a)].
For each odd $N_\mr{occ}$, the charge densities fluctuates across sites, averaging around $N_\mr{occ}/6$.
Site-dependent fluctuations in charge density arise due to electron interactions and quantum effects. 
However, these fluctuations exhibit a repeating pattern beyond three atomic sites, reflecting the
underlying symmetry of the system.
Additionally, the difference between the charge densities for successive odd occupation numbers remains 
nearly constant. 
This difference is approximately $ \approx 0.33$, which corresponds to the average charge 
density of an electron pair ($\uparrow \downarrow$) uniformly distributed across 
the six atomic sites. 
This suggests a systematic charge accumulation as the electron number increases, maintaining a balanced
distribution within the lattice.

The corresponding spin densities for odd $N_\mr{occ}$ [Fig.~\ref{fig6}(b)] alternate significantly
between lattice sites, forming a staggered pattern.
The alternating positive and negative spin densities suggest an antiferromagnetic-like arrangement, 
where neighboring sites preferentially adopt opposite spin orientations.
In contrast to the smoothly varying charge densities, spin densities exhibit sharper oscillations with
clear site-to-site alternation.
This behavior is characteristic of correlated systems, where electrons tend to avoid double occupancy 
to minimize interaction energy, leading to pronounced spin polarization.
Notably, the spin densities for odd occupation numbers exhibit mirrored patterns.
This symmetry arises because of particle-hole conjugation and the intrinsic properties of the Hubbard
Hamiltonian on a bipartite lattice.

For fully paired electron configurations, i.e., even $N_\mr{occ}$, 
the charge densities are uniformly distributed across the atomic sites, Fig.~\ref{fig6}(a).
This suggests that, in the absence of unpaired electrons, the charge distribution becomes more homogeneous,
reducing strong fluctuations seen in odd occupation numbers.
The constant charge density for even $N_\mr{occ}$ suggests uniform electron pairing, characteristic
of a non-magnetic, correlated metallic state.
As seen, the corresponding spin densities in these cases are zero at each atomic site [Fig.~\ref{fig6}(b)],
consistent with singlet-paired electrons forming a spin-balanced state.
It is worth noting that we omit the adiabatic simulation results for charge and spin
densities, as well as charge and spin correlations, at $N_\mr{occ} = 4$ and $8$ in
Fig.~\ref{fig6}, due to the vanishing energy gap
$\Delta_\mr{min} (\eta)|_{\eta \rightarrow 0} \rightarrow 0$, as defined in Eq.~\eqref{eq:T_evol}. 
This zero gap prevents the adiabatic evolution from reaching the true ground state within a
practical time frame for these two configurations.

To complete our investigation, we now examine the charge correlations $C_{0,j}^c$ and spin
correlations $C_{0,j}^s$ for different values of 
$N_\mr{occ}$ as depicted in Figs.~\ref{fig6}(c) and \ref{fig6}(d), respectively.
Due to particle-hole symmetry between states below and above half-filling, we present results only for
$N\mr{occ} = 2,3, \ldots, 6$.
As seen in Fig.~\ref{fig6}(c), the charge correlations are negative across all sites and $N_\mr{occ}$.
This indicates anti-correlated charge fluctuations: if site $j=0$ has a higher-than-average electron
density, site $j$ tends to have a lower-than-average density, driven by the repulsive interaction 
$U_0$.
For all $N_\mr{occ}$, the charge correlations at sites $j=1$ and $j=5$ are the most negative, indicating 
strong anti-correlation with site $j=0$.
In a hexagonal geometry [Fig.~\ref{fig1}(a)], sites $j=1$ and $j=5$ are nearest neighbors to site $j=0$, suggesting
that charge anti-correlations are strongest at these distances.
For sites $j=2,3,4$, weaker anti-correlations are observed, as they are farther from site $j=0$.
A symmetric trend in charge correlations is evident for all $N_\mr{occ}$, consistent with 
the underlying hexagonal structure.
However, this symmetry is exact for even $N_\mr{occ}$ and slightly broken for odd $N_\mr{occ}$, likely due to 
an influence from the fermionic parity or interaction effects in the system.
For $N\mr{occ}=6$ (purple), the system is at half-filling, where electron repulsion is maximized. 
This leads to strongest anti-correlations, as electrons experience enhanced localization due to Coulomb
interactions. 
This behavior is characteristic of a Mott-like insulating state, where charge movement is suppressed, 
and each site tends to be occupied by one electron on average.
For $N\mr{occ}=2$ (blue), the system exhibits the weakest charge anti-correlations since it is far from 
half-filling. 
In this regime, electrons have more available empty sites, allowing them to easily avoid each other, 
reducing charge correlations.

Figure \ref{fig6}(d) presents the corresponding spin correlations for the charge correlations shown 
in Fig.~\ref{fig6}(a).
The evolution of spin correlations with increasing electron occupation number $N_\mr{occ}$
reveals key signatures of electronic interactions and magnetic ordering in the moderately correlated 
Hubbard model with $U_0=3$ and $t=1$.
At low filling, spin correlations remain weakly negative ($N_\mr{occ}=2,3$), suggesting a weak tendency 
toward antiferromagnetism (AFM) due to the lack of a well-defined spin structure.
As $N_\mr{occ}$ increases, spin correlations transition from weak, uniformly negative values at low 
filling to a stronger alternating pattern at half-filling. 
This progression reflects the increasing influence of e-e interactions, leading to the emergence of
short-range AFM. 
At $N_\mr{occ}=6$ (half-filling), the system exhibits its strongest AFM correlations, characteristic of 
a Mott-like insulating state where Coulomb repulsion dominates charge fluctuations.

Finally, we conclude this section by briefly discussing the IQPE algorithm and adiabatic quantum 
simulation in simulating the Hubbard model on a six-site graphene hexagon.
As demonstrated, the strong quantitative agreement between exact diagonalization
and noiseless IQPE simulations highlights the accuracy and reliability of the IQPE method in capturing
GSE and, more generally, the eigenvalues of strongly correlated systems. 
In the presence of interactions, IQPE remains highly accurate, even when using SDs as initial states.
On the other hand, adiabatic state preparation proves to be an efficient approach for computing local
observables, such as charge and spin densities, as well as charge and spin correlations. 
However, the time evolution operator in Eq.~\eqref{eq:T_evol} plays a crucial role in the success 
of quantum simulations. 
Notably, for $N_\mr{occ} = 4$ and $8$, the algorithm fails to converge to the exact GSE due to the behavior 
of the minimum energy gap $\Delta_\mr{min} (\eta)|_{\eta \rightarrow 0}$, which leads to an infinitely 
long evolution time--a fundamental limitation of adiabatic methods in such cases.

\subsection{Noisy simulation and experiment} \label{ss_noisy}
Studying the Hubbard model on real quantum hardware is essential for evaluating the practical
feasibility of quantum algorithms like IQPE and adiabatic quantum simulation under realistic
conditions, where gate errors, decoherence, and qubit connectivity constraints can significantly
degrade performance compared to noiseless simulations that confirm their accuracy. 
To systematically analyze these effects, we construct a noise model tailored to \ibmstr's
properties, incorporating three dominant noise channels--depolarizing errors, thermal relaxation 
errors, and readout errors--reflecting the primary error mechanisms in superconducting qubit systems,
and use it to investigate how noise, error rates, and circuit depth limitations impact the accuracy
and reliability of IQPE in estimating GSEs of quantum systems.

\begin{figure*}
	\centering
	\includegraphics[width = 17.5 cm]{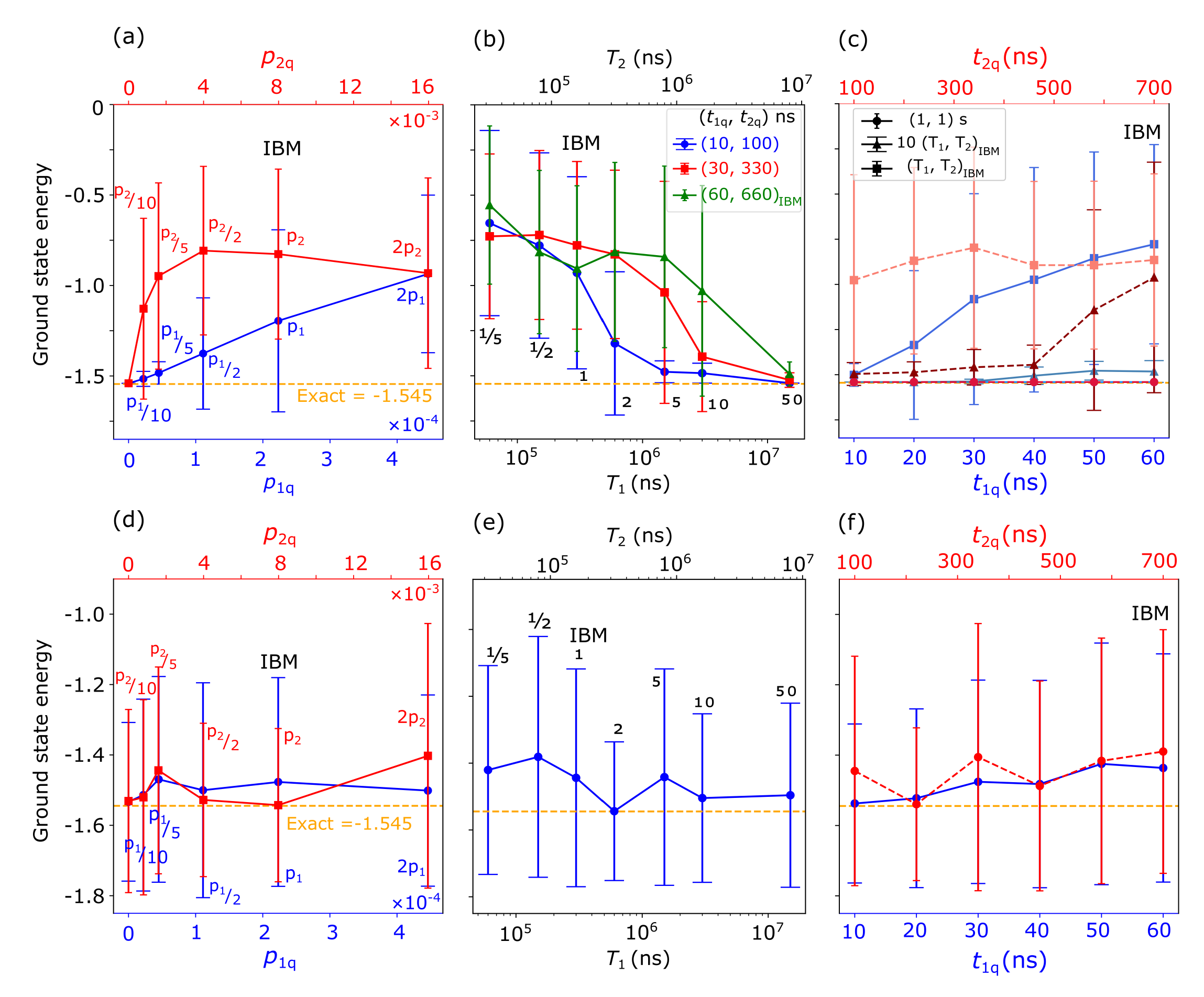}
	\caption{The upper panels show the effect of each noise parameter on IQPE estimation of 
		the GSE for a half-filled three-atomic-site Hubbard model, assuming all other noise
		channels are absent or set to their ideal values.
		(a) Effect of depolarizing errors -- Mean GSE (solid lines) and its variance 
		(shaded regions, representing standard deviation across $50$ runs with $50,000$ shots each) 
		are plotted as functions 
		of single-qubit ($p_{1q}$, lower axis) and two-qubit ($p_{2q}$, upper axis) 
		depolarizing errors, expressed as factors 
		of \ibmstr's median gate errors ($p_1$ and $p_2$). 
		The exact GSE ($-1.545$) is shown by the orange dashed line. 
		(b) Effect of thermal relaxation times --
		Mean GSE and variance as a function of $T_1$ (lower axis) and $T_2$ (upper axis) for different
		gate duration pairs ($t_{1q},\ t_{2q}$): ($10, 100$) ns (blue circles), ($30, 330$) ns 
		(red squares), and ($460, 660$) ns (green triangles). 
		The results are presented for a range of ($T_1$, $T_2$) values, each scaled 
		from the  \ibmstr\ baseline by factors of $[1/5, 1/2, 1, 2, 5, 10, 50]$.
		(c)	Effect of gate durations --
		Mean GSE and variance as a function of single-qubit gate duration ($t_{1q}$, lower axis, blue)
		and two-qubit gate duration ($t_{2q}$, upper axis, red) for three sets of thermal relaxation
		parameters: ($T_1,\ T_2$) = ($1,\ 1$) s (ideal, circles), $10 \times (T_1,\ T_2)_\text{IBM}$ 
		(triangles), and $(T_1,\ T_2)_\text{IBM}$ (squares). 
		Bluish solid lines represent $t_{1q}$ variations, while reddish dashed lines represent 
		$t_{2q}$ variations.
		The lower panels illustrate the impact of (d) depolarizing errors, (e) thermal 
		relaxation times, and (f) gate durations on IQPE GSE estimation, each evaluated 
		while keeping the other noise parameters fixed at \ibmstr's baseline values.
	}
	\label{fig7}
\end{figure*}
Depolarizing errors (gate infidelity) model the infidelity of quantum gate operations, capturing the 
random perturbations that occur during gate execution. 
In our noise model, single-qubit gates (e.g., $X$) and two-qubit gates (e.g., ECR) 
are assigned depolarizing channels with error probabilities derived from \ibmstr's median 
gate errors: $p_1 = 2.230 \times 10^{-4}$ for sing-qubit gate and $p_2 = 7.986 \times 10^{-3}$ for
two-qubit gates. 
This noise type simulates a process where, with a given probability, the qubit's state is replaced by 
a completely mixed state, effectively applying Pauli $X$, $Y$, or $Z$ errors with equal likelihood.
Depolarizing errors are particularly significant for two-qubit gates, which exhibit higher 
error rates due to their longer duration and complexity. 
In the context of IQPE, these errors may degrade the precision of controlled operations, potentially 
leading to phase estimation inaccuracies.

Thermal relaxation errors account for the decoherence that qubits experience due to interactions 
with their environment, both during active gate operations and idle periods. 
Parameterized by the hardware's $T_1$ (energy relaxation time, $\sim 300~\mu$s) and $T_2$ 
(dephasing time, $\sim 160~\mu$s), this noise channel models two processes: relaxation, where an excited qubit
($|1 \rangle$) decays to the ground state ($|0 \rangle$), and dephasing, where phase coherence in
superpositions is lost. 
We incorporate gate durations, $t_1$ ($\sim 60$ ns) for single-qubit gates and $t_2$ 
($\sim 660$ ns) for two-qubit gates, into the thermal relaxation model to simulate 
idle noise on inactive qubits, reflecting realistic circuit scheduling. 
For IQPE, which relies on precise phase accumulation over multiple iterations, thermal relaxation 
could introduce cumulative errors, especially in deeper circuits where idle times amplify decoherence effects.

Readout errors represent the misclassification of qubit states during measurement, a critical step in
extracting results from quantum algorithms like IQPE. 
Using \ibmstr's median readout error of $1.953 \times 10^{-2}$, we model this noise as an
asymmetric error channel with probabilities $P_{0|1} =  0.016$ and $P_{1|0} = 0.022$. 
This error type reflects the hardware's difficulty in distinguishing $|0 \rangle$ from $|1 \rangle$, 
often due to noise in the measurement apparatus or qubit relaxation during the measurement process 
(typically $1\ \mu$s). 

In the noiseless case, we modeled the full hexagonal ring with all six atomic sites to precisely
capture the complete lattice connectivity and electronic interactions of the graphene structure. 
For the noisy simulations emulating IBM QPUs, however, we reduce the system register to $N = 3$ 
atomic sites. 
This reduction in qubit count ensures simulations remain fast and feasible, enabling a detailed
investigation of noise effects (depolarizing errors, thermal relaxation, and readout imperfections) on 
IQPE performance.
The simulations remain calibrated to hardware-realistic parameters while still capturing the 
essential features of the graphene Hubbard model.

First, we analyze the impact of each noise type individually--excluding the influence of other errors--on 
the energy estimation accuracy of the IQPE algorithm.
To do so, we isolate each noise parameter, vary it over a broad range, and plot the mean GSE 
(averaged over 50 runs with $50,000$ shots each) along with its variance to assess improvements 
in accuracy and reliability.
For this analysis, we focus on the GSE of a half-filled three-atomic-site structure at $U_0=3$.

Figure \ref{fig7}(a) depicts the GSE as a function of single-qubit ($p_{1q}$, blue) and two-qubit 
($p_{2q}$, red) depolarizing error rates.
The result is expressed as the factor of \ibmstr's median gate errors $p_1$ and $p_2$.
The exact GSE is $-1.545$, indicated by the orange dashed line.
For small error rates ($p_{1q} = p_1/10, p_1/5$), the estimated GSE remains close to the exact value,
with minimal variance, indicating that the algorithm is robust to small single-qubit gate errors.
As $p_{1q}$ increases, the estimated GSE deviates from the exact value, showing a smooth, monotonic
increase with larger variance. 
In contrast, two-qubit errors have a markedly more disruptive impact--even at low error
rates--shifting the estimated GSE to higher values and causing a sharp increase in variance.
This behavior highlights the critical importance of two-qubit gate fidelity for achieving accurate 
GSE estimates in NISQ-era simulations of quantum many-body systems like the Hubbard model, 
where multi-qubit interactions are essential for capturing complex electron correlations.

Figure \ref{fig7}(b) illustrates the mean estimated GSE as a function of thermal relaxation 
times ($T_1, T_2$) for different single- and two-qubit gate durations ($t_{1q}, t_{2q}$), 
as indicated in the figure.
The GSE is presented for a range of ($T_1, T_2$) values, each scaled from the baseline values 
of the \ibmstr\ quantum processor by factors of $[1/5, 1/2, 1, 2, 5, 10, 50]$.
For short coherence times, below the IBM baseline values, all gate durations result in 
substantially overestimated GSEs, reflecting significant decoherence effects. 
Beyond the baseline values, the GSE for the shortest gate durations (blue) drops sharply and
converges towards the exact GSE (orange dashed line) as the $T_1$ and $T_2$ increase, reflecting
reduced error rates.
In contrast, the intermediate (red) and current IBM (green) gate durations yield progressively
higher GSE estimates up to approximately 5 times the IBM baseline for $T_1$ and $T_2$, indicating
increased sensitivity to thermal noise at these scales.
However, beyond this point, the GSE estimates for these longer gate durations gradually decrease,
eventually converging to the exact GSE at around 50 times the baseline $(T_1, T_2)$ values.
This suggests a threshold behavior, where the coherence times become sufficiently long to suppress
the dominant relaxation errors, resulting in accurate GSE estimation regardless of gate duration.
Note that the shortest gate durations result in high variance up to approximately 2 times the IBM
baseline $(T_1, T_2)$ values. 
Beyond this point, the variance decreases significantly as coherence times improve, indicating 
more consistent and accurate GSE estimation.
For the intermediate and IBM gate durations, a substantial reduction in variance is also observed
beyond 10 times the $(T_1, T_2)$ baseline values.

Shown in Fig.~\ref{fig7}(c) illustrates the estimated GSE as a function of single-qubit
($t_{1q}$, lower axis, blue) and two-qubit ($t_{2q}$, upper axis, red) gate durations for three
different thermal relaxation parameter sets ($T_1,\ T_2$): ($1,\ 1$) s (circles), representing
effectively error-free coherence times, $10 \times (T_1,\ T_2)_\text{IBM}$ (triangles), representing 
tenfold enhanced IBM coherence times, and $(T_1,\ T_2)_\text{IBM}$ (squares), representing 
typical IBM coherence times.
Bluish solid lines represent $t_{1q}$ variations, while reddish dashed lines represent $t_{2q}$
variations.
For the ideal case, the GSE estimates remain relatively stable and close to the exact value,
across the full range of gate durations, reflecting minimal decoherence effects.
At $10 \times (T_1,\ T_2)_\text{IBM}$, the GSE remains relatively stable over the full range
of $t_{1q}$ values (bluish triangles), showing only minor deviations from the exact GSE,
reflecting the robustness of single-qubit operations at extended coherence times. 
However, as a function of $t_{2q}$ (reddish triangles), the GSE exhibits significant overestimation
and high variance beyond approximately $t_{2q} = 450$ ns, highlighting the cumulative impact of 
longer two-qubit gate durations on phase fidelity.
For the IBM baseline coherence times $(T1,\ T2)_\text{IBM}$ (blue, squares), the GSE increases
gradually from near the exact value at the shortest $t_{1q} = 10$ ns to approximately $-0.8$ at 
$t_{1q} = 60$~ns, reflecting the increasing phase error with longer gate durations.
As a function of $t_{2q}$ (reddish squares), the GSE exhibits a more pronounced degradation, 
dropping to approximately $-1.0$ at the shortest $t_{2q}$ values and showing a gradual 
recovery with increasing $t_{2q}$.
This trend indicates the greater vulnerability of multi-qubit gates to thermal relaxation effects,
where the cumulative impact of error channels can significantly degrade the overall phase coherence,
leading to overestimated GSE values.
Our investigation (not shown here) shows that the IQPE algorithm is robust to a wide range of 
readout errors $[0, 0.2]$.

In the lower panel of Fig.~\ref{fig7}, we investigate the impact of each noise on IQPE GSE estimation, 
while keeping the other parameters fixed at \ibmstr's baseline values, reflecting a more realistic 
device noise environment.
In Fig.~\ref{fig7}(d), the GSE estimates remain closer to the exact value across 
a wider range of $p_{1q}$ and $p_{2q}$ values compared to the isolated depolarizing error cases
[Fig.~\ref{fig7}(a)], indicating a partial compensation from other noise sources.
However, the GSE still drifts upward at higher error rates, reflecting the cumulative effect of
multi-channel noise.
The error bars reflect high variance for all range of $p_{1q}$ and $p_{2q}$ values, indicating significant
uncertainty in GSE estimates.
Compared to the isolated noise plots, the GSE here shows less severe overestimation for similar 
$p_{1q}$ and $p_{2q}$ ranges, emphasizing the complex interplay between multiple noise sources in 
practical NISQ devices.

Figure \ref{fig7}(e) presents the mean estimated GSE as a function of thermal relaxation times 
($T_1,\ T_2$), with other noise parameters fixed at typical IBM values. 
The data are plotted for various $T_1, T_2$, scaling factors relative to the IBM baseline as 
shown in the plot.
Unlike the isolated thermal relaxation case [Fig.~\ref{fig7}(b)], the GSE in Fig.~\ref{fig7}(e),
remains relatively stable across the entire range of $T_1$ and $T_2$ values, with a mean close 
to the exact value.
Despite the stable mean, the error bars remain consistently large across the range, reflecting significant
uncertainty in GSE estimates.

In Fig.~\ref{fig7}(f), the estimated GSE exhibits an oscillatory, linear-like overestimation as 
a function of both $t_{1q}$ and $t_{2q}$, with the effect being more pronounced for the 
two-qubit gate durations. 
The larger amplitude of oscillations observed along the $t_{2q}$ axis further underscores the
heightened sensitivity of two-qubit gates to duration-induced decoherence, consistent with their more
complex entangling nature and higher error rates.

The collective analysis of the noise plots reveals a critical insight into the interplay 
between various quantum noise sources and their impact on the GSE estimation using the 
IQPE algorithm. 
While single-qubit depolarizing errors ($p_{1q}$) generally induce smaller, more controlled deviations 
in the GSE, two-qubit errors ($p_{2q}$) exhibit a substantially larger and less predictable impact,
reflecting the higher sensitivity of entangling gates to noise. 
The thermal relaxation times ($T_1,\ T_2$) also play an axial role, where longer coherence times
significantly suppress energy overestimations, but only if the gate durations remain sufficiently short.
This points to the importance of optimizing both coherence and gate fidelity in reducing energy errors.
Taken together, these findings highlight the need for careful noise mitigation strategies in near-term
quantum devices, particularly emphasizing the reduction of two-qubit error rates and the extension of
coherence times to achieve reliable GSE estimates for complex quantum systems like the Hubbard model.

\begin{figure}
	\centering
	\includegraphics[width = 8 cm]{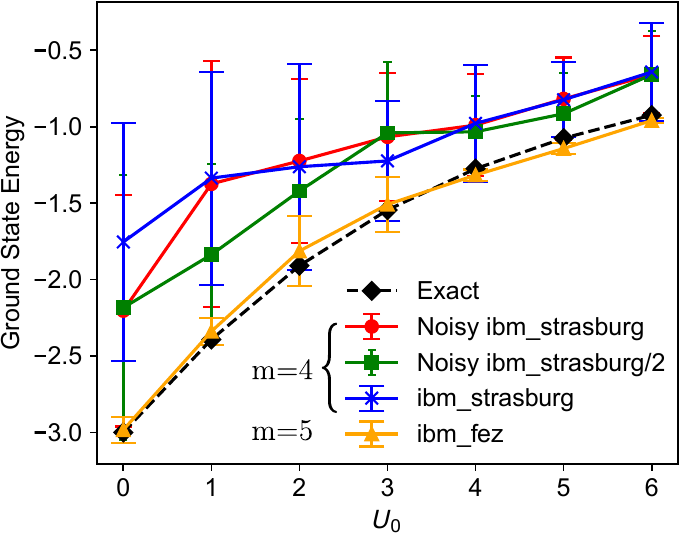}
	\caption{GSE as a function of on-site interaction strength $U_0$ for a half-filled
		three-atomic-site Hubbard model. 
		The plot compares results from exact diagonalization (black diamonds, dashed line),
		experimental runs on \ibmstr\ quantum hardware (blue ``$\times$"), noisy simulations 
		using \ibmstr-calibrated noise parameters (red circles), and simulations with noise
		parameters reduced by a factor of 2 (green squares).		
		All results use $m = 4$ and $N_\text{trot} = 12$, with error bars representing the
		standard deviation across 20 runs of 10,000 shots each. 
		Experimental GSEs obtained on \ibmfez\ with improved parameters 
		($m = 5$, $N_\text{trot} = 15$) are represented by orange triangles.
	}
	\label{fig8}
\end{figure}

Finally, we execute the IQPE algorithm on the IBM's real quantum hardware, \ibmstr, and obtained the GSE for
a half-filled three-atomic-site structure as a function of the on-site interaction $U_0$, as shown in
Fig.~\ref{fig8}.
Due to current hardware constraints, the maximum binary precision achievable on \ibmstr\ in our experiments was $m = 4$, with a Trotter step number of $N_\mathrm{trot} = 12$.
We also include results from noise simulations based on \ibmstr's hardware-calibrated noise parameters, 
along with simulations using noise parameters reduced by a factor of two, while keeping the same 
binary precision and Trotter step number.
In all cases, we averaged the results over 20 runs with 50,000 shots each.  
The dashed line represents the exact diagonalization results.
Experimental data from IBM quantum hardware (blue ``$\times$") closely follow the noisy simulation results
with IBM-calibrated noise parameters (red circles), particularly for higher $U_0$ values, although 
both consistently overestimate the GSE compared to the exact solution due to accumulated noise effects. 
Notably, the green squares, representing simulations with improved (halved) noise levels, show a 
clearer approach to the exact curve, in the weakly interacting regime ($U_0 \lessapprox 3$), 
suggesting that better hardware fidelity can improve quantum simulation accuracy.
Across all cases, the error bars indicate remarkable variance at small $U_0$, highlighting greater
sensitivity to noise in the weakly correlated regime.
As seen, however, all cases closely follow each other with reduced variance at $U_0 \gtrapprox 3$,
indicating that the impact of noise diminishes in the strong interaction regime.
The difference in noise sensitivity across $U_0$ can be attributed to the circuit complexity tied 
to the Hamiltonian structure. 
In the kinetic-dominated regime, non-commuting terms lead to deeper, 
noisier circuits, whereas in the interaction-dominated regime, the commuting, diagonal nature 
of the interaction terms allows for more noise-resilient simulations.
However, implementing the IQPE algorithm on \ibmfez\ with improved run parameters, $m = 5$ and $N_\mathrm{trot} = 15$ (averaged over 5 runs with 50,000 shots each), yields GSE values closely matching the exact result, as shown by the orange solid line in Fig.~\ref{fig8}.
This result highlights the influence of hardware noise and the importance of parameter optimization for accurate IQPE-based energy estimation.

\section{Conclusion and Remarks} \label{con}
In this work, we investigated the ground state properties of the Hubbard model on a six-site graphene
hexagon using quantum simulation techniques. 
The study leveraged the IQPE algorithm and adiabatic quantum simulation, implemented via noiseless,
noisy, and real quantum hardware simulations.
As the system maps to a 1D Hubbard chain with PBC, we also showed that the non-local JW string simplifies to a global phase factor, which greatly reduces circuit depth and noise.

The noiseless simulations confirmed the capability of the IQPE algorithm to accurately reproduce the 
exact GSE (within a few iterations), demonstrating its effectiveness for correlated fermionic systems such as the Hubbard model.
Notably, our investigation revealed that initializing the IQPE with a single SD is
sufficient to recover the correct ground state eigenvalue in small-scale Hubbard systems.
Furthermore, to probe local observables such as charge and spin correlations, we employed adiabatic 
quantum simulation, which yielded results in excellent agreement with those obtained from exact
diagonalization.

To investigate noise-related limitations in IQPE algorithm, we performed noisy simulations in two scenarios.
First, we evaluated the impact of each noise type in isolation, excluding all other errors, to assess 
their individual effects on IQPE energy estimation. 
Second, we studied each noise type while keeping the others fixed at IBM's (\ibmstr) baseline values, reflecting 
a more realistic hardware noise environment. 
To reduce computational cost, the system size was limited to $N=3$ atomic sites.

In the idealized context, we found thermal relaxation and two-qubit gate errors to be the most detrimental, 
causing significant energy overestimations and high variances.
Longer gate durations also degraded performance, particularly for two-qubit operations, highlighting
their critical role in IQPE accuracy.
In contrast, readout errors were found to have a negligible impact on the overall accuracy within 
the scope of our simulations.
Compared to the full-noise simulations, these idealized studies revealed more systematic and
interpretable trends, offering clearer diagnostics for performance bottlenecks. 
They highlight that even in the absence of compounding errors, individual noise sources--particularly
two-qubit gate imperfections and short coherence times--pose fundamental challenges to accurate 
quantum simulation in the NISQ regime.

In the second scenario, the GSE shows less severe overestimation across a wider range of studied noise
parameter values compared to the isolated cases, indicating a partial compensation from other 
noise sources.
In this case, our analysis showed that two-qubit gate errors and durations were the most harmful 
and unpredictable, significantly shifting GSE estimates and increasing variance.
Regarding thermal relaxation effects, increasing $T_1$ and $T_2$ coherence times generally improved GSE
accuracy. 
However, the benefits were more pronounced when gate durations were short, highlighting the interplay
between coherent control and decoherence timescales.

Finally, execution on IBM quantum hardware confirmed that IQPE can accurately estimate GSEs, though its performance is limited by hardware noise and circuit depth. Using the \ibmstr\ and \ibmfez\ devices for a reduced three-site system, we found that \ibmfez\ was able to run with improved tuning parameters, yielding results that closely match the exact values.
This work highlights both the potential and current challenges of quantum computing for materials simulation, providing a benchmark for IQPE performance on strongly correlated systems under realistic noise.

\renewcommand{\theequation}{A\arabic{equation}}
\setcounter{equation}{0}
\renewcommand{\thefigure}{A\arabic{figure}}
\setcounter{figure}{0}
\setcounter{section}{0}
\begin{figure*} [htbp]
	\centering
	\includegraphics[width=17cm]{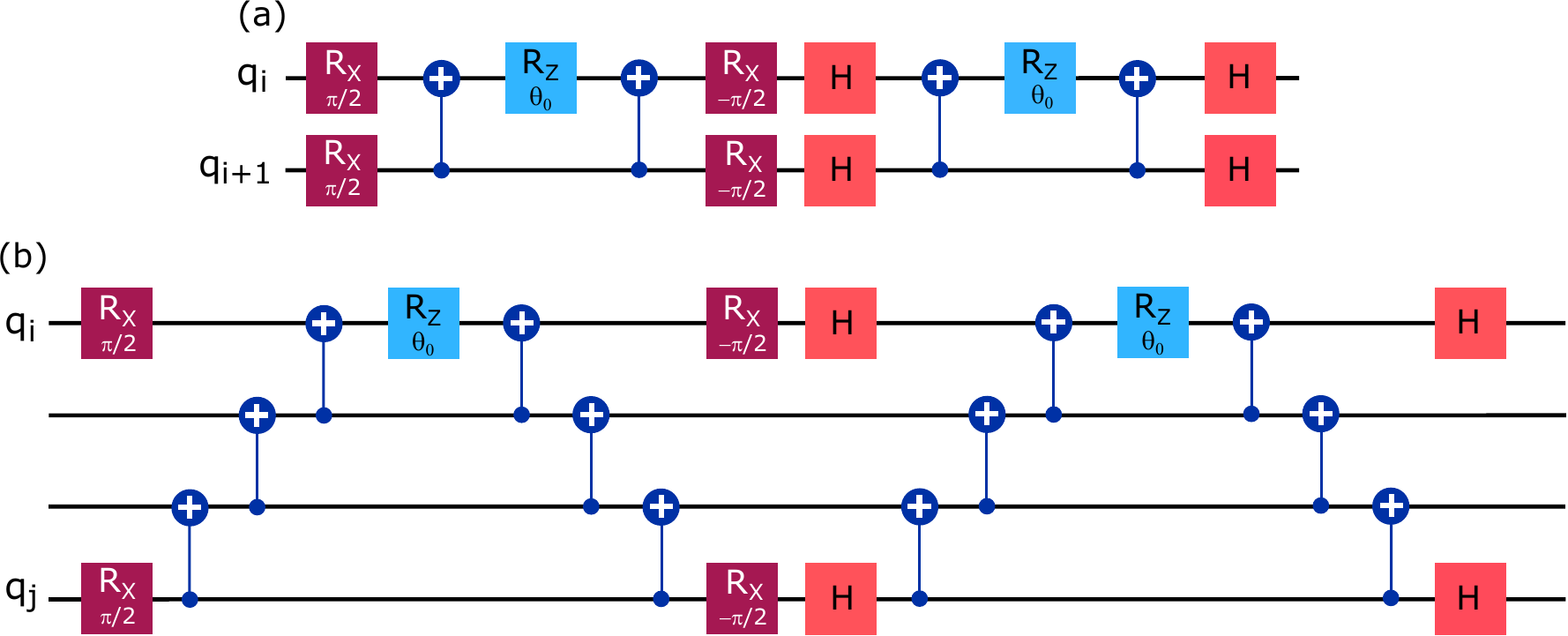}
	\caption{Quantum circuit to implement time evolution under the hopping term 
		$h_0 = -\gamma_0 (a^\dagger_{i \s} a_{j \s} + a^\dagger_{j \s} a_{i \s})$.
		The time step is $\theta_0 = -\gamma_0 t$.
		Panel (a) shows the quantum circuit corresponding to the hopping term for two adjacent atomic 
		sites labeled sequentially as $i$ and $i+1$.
		Panel (b) depicts the circuit for two adjacent sites, with non-sequential labeling.
	}
	\label{figA1}
\end{figure*}

\begin{figure} [htbp]
	\centering
	\includegraphics[width=8cm]{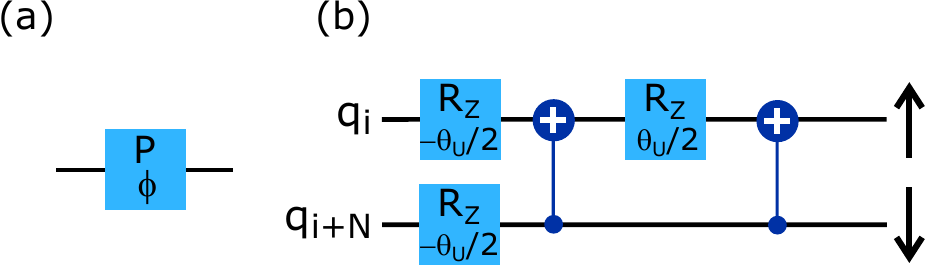}
	\caption{Quantum circuits to implement time evolution for (a) the number operator  
		$n_{i \s} = a^\dagger_{i \s} a_{i \s}$ and (b) on-site interaction term $U_0 n_{i\s} n_{j\s'}$.
		In panel (a), $P$ denotes a Phase gate with $\phi = -t$, while in panel (b), the time step 
		is given by $\theta_U = U_0 t$.
	}
	\label{figA2}
\end{figure}

\section*{Acknowledgments}
This work is supported by the National Research Foundation
of Korea (grant numbers NRF-2022M3K2A108385811, RS-2023-00257561).

\section*{Appendix: Circuits for the Hubbard Hamiltonian} \label{append}

\textbf{Quantum circuits:} The unitary evolution under the hopping term $h_0$, represented in Eq.~\eqref{eq:hop} in the main text,
is expressed as
\begin{equation}
	\mc U_0(t) = \exp(-\rmi h_0 t) = \exp\Big[-\rmi \frac{\theta_0}{2}(X_i X_j + Y_i Y_j) Z_\mr{JW} \Big],
\end{equation}
whose quantum circuit is illustrated in Fig.~\ref{figA1}.
Figure \ref{figA1}(a) shows the circuit for two adjacent atomic sites, labeled 
sequentially as $i$ and $i+1$, while Fig.~\ref{figA1}(b) depicts the circuit for two adjacent 
atomic sites with non-sequential labeling.
The time step is $\theta_0 = -\gamma_0\, t$.

The quantum circuit for the unitary evolution under the number operator 
(Eq.~\eqref{eq:num} in the main text)
\begin{equation}
	\mc U_n(t) = \exp(-\rmi a^\dagger_{i} a_{i} t) = 
	\exp\big[-\rmi \frac{1}{2}(I_i - Z_i) t\big],
\end{equation}
is shown in Fig.~\ref{figA2}(a), which can be considered as a Phase gate defied as 
$
P(\phi) = \begin{bmatrix}
	1 & 0 \\
	0 & \e^{\rmi \phi}
\end{bmatrix}
$
with the phase shift of $\phi = - t$.

The on-site interaction term is expressed in Eq.~\eqref{eq:hub}, and its unitary evolution is given by 
\begin{multline}
	\mc U_U(t) = \exp(-\rmi U_0 n_{i} n_{j} t) \\
	= \exp\Big[-\rmi \frac{\theta_U}{4}(I_i I_j - I_i Z_j - Z_i I_j + Z_i Z_j)\Big],
\end{multline}
with the corresponding quantum circuit depicted in Fig.~\ref{figA2}(b).
Here, the time step is $\theta_U = U_0 t$.
Note that due to the commutativity of interaction terms in the Hubbard model, they can be applied
simultaneously.

\textbf{Proof of Eq.~\eqref{eq:jw_per}:} Here, we provide a detailed derivation of Eq.~\eqref{eq:jw_per} in the main text, demonstrating the simplification of the JW string for a Hubbard chain with PBC.
We begin by recalling the JW transformation for spinless fermions:
\begin{align} \label{eq:JW}
	a_j &= (\otimes_{k=1}^{j-1} Z_k) \otimes  \sigma_j, \\
	a_j^\dagger &= (\otimes_{k=1}^{j-1} Z_k) \otimes \sigma_j^\dagger,
\end{align}
where $\sigma_j = \frac{1}{2}(X_j + \rmi Y_j)$ and $\sigma_j^\dagger = \frac{1}{2}(X_j - \rmi Y_j)$ are the spin lowering and raising operators at site $j$, respectively.
We now consider the hopping term between the last and first sites, corresponding to the PBC:
\begin{align} \label{eq:a_dag-a_pbc}
	a_N^\dagger a_1 = \left(\otimes_{k=1}^{N-1} Z_k\right) \otimes \sigma_N^\dagger \sigma_1 = \sigma_1^z \sigma_1 \left(\otimes_{k=2}^{N-1} Z_k\right) \otimes \sigma_N^\dagger.
\end{align}
Noting that $\sigma_1^z \sigma_1 = \sigma_1$, we can rewrite Eq. \eqref{eq:a_dag-a_pbc} as:
\begin{align}
	a_N^\dagger a_1 = \sigma_1 \left(\otimes_{k=2}^{N-1} Z_k\right) \otimes \sigma_N^\dagger.
\end{align}
Using the number operator identity $Z_k = I_k - 2n_k$ [Eq.\ \eqref{eq:num}], the remaining JW string becomes:
\begin{equation}
	\otimes_{k=2}^{N-1} Z_k = \otimes_{k=2}^{N-1} ( I_k - 2n_k)
	= (-1)^{\sum_{k=2}^{N-1} n_k} \otimes_{k=2}^{N-1} I_k.
\end{equation}
Defining the total fermion number in the spin sector as $N_f = \sum_{k=1}^N n_k$, we find
$\sum_{k=2}^{N-1} n_k = N_f - n_1 - n_N$, which leads to:
\begin{equation}  \label{eq:Z_jw_sim}
	\otimes_{k=2}^{N-1} Z_k
	= (-1)^{N_f - n_N -n_1} \otimes_{k=2}^{N-1} I_k.
\end{equation}
Substituting Eq.~\eqref{eq:Z_jw_sim} back, we obtain:
\begin{align} 
	a_N^\dagger a_1 =  (-1)^{N_f - n_N -n_1} \sigma_1 \otimes_{k=2}^{N-1} I_k \otimes \sigma_N^\dagger.
\end{align}
Physically, $a_N^\dagger a_1$ describes a fermion hopping from site 1 to site $N$. 
For this process to occur, site 1 must be occupied ($n_1 = 1$) and site $N$ must be empty 
($n_N = 0$), yielding:
\begin{equation}
	a_N^\dagger a_1 \rightarrow (-1)^{N_f-1}  \sigma_1 \otimes_{k=2}^{N-1} I_k 
	\otimes \sigma_N^\dagger.
\end{equation}
This result confirms the simplification presented in Eq.~\eqref{eq:jw_per} of the main text, showing that the non-local JW string under PBC reduces to a global phase factor depending on the total fermion number in the fixed-occupancy sector.
\\

\textbf{Ethics Approval declaration:}
Ethics approval is not required for this study as it did not involve human or animal subjects.
   


\begin{thebibliography}{99}

\bibitem{Georgescu2014} 
I. M. Georgescu, S. Ashhab, and F. Nori, 
Quantum simulation, 
Rev. Mod. Phys. \textbf{86}, 153 (2014).

\bibitem{Bauer2016} 
B. Bauer, D. Wecker, A.J. Millis, M. B. Hastings, and M. Troyer,
Hybrid quantum-classical approach to correlated materials,
Phys. Rev. X \textbf{6}, 031045 (2016).

\bibitem{Bauer2020}
B. Bauer, S. Bravyi, M. Motta, and G. K.-L. Chan, 
Quantum algorithms for quantum chemistry and quantum materials science,
Chem. Rev. \textbf{120}, 12685 (2020).

\bibitem{Hubbard1963} 
J. Hubbard, 
Electron correlations in narrow energy bands,
Proc. R. Soc. A \textbf{276}, 238 (1963).

\bibitem{Essler2005} 
F. H. L. Essler, H. Frahm, F. G\"{o}hmann, A. Kl\"{u}mper, and V. E. Korepin, 
\textit{The one-dimensional Hubbard model}, 
(Cambridge University Press, 2005).

\bibitem{Scalapino2007} 
D. J. Scalapino, ``Numerical studies of the 2D Hubbard model"
In \textit{Handbook of High-Temperature Superconductivity: Theory and Experiment}, 
(Springer, New York, 2007), pp. 495–526.

\bibitem{Scalapino2012} 
D. J. Scalapino,
A common thread: The pairing interaction for unconventional superconductors,
Rev. Mod. Phys. \textbf{84}, 1383 (2012).

\bibitem{LeBlanc2015}
J. P. LeBlanc, A. E. Antipov, F. Becca, I. W. Bulik, G. K.-L. Chan, C.-M. Chung, Y. Deng,
M. Ferrero, T. M. Henderson \textit{et al}.,
Solutions of the two-dimensional Hubbard model: benchmarks and results from a wide range of
numerical algorithms,
Phys. Rev. X \textbf{5}, 041041 (2015).

\bibitem{Qin2022}
M. Qin, T. Sch\"{a}fer, S. Andergassen, P. Corboz, and E. Gull, 
The Hubbard model: A computational perspective,
Annu. Rev. Condens. Matter Phys. \textbf{13}, 275 (2022).

\bibitem{Mott1949} 
N. F. Mott, 
The basis of the electron theory of metals, with special reference to the transition metals,
Proc. Phys. Soc. A \textbf{62}, 416 (1949).
 
\bibitem{Lee2008} 
P. A. Lee,
From high temperature superconductivity to quantum spin liquid: progress in strong correlation physics,
Rep. Prog. Phys. \textbf{71}, 012501 (2008).

\bibitem{Yamada2005} 
S. Yamada, T. Imamura, and M. Machida, 
16.447 TFlops and 159-Billion-dimensional exact-diagonalization for trapped 
Fermion-Hubbard model on the earth simulator,
SC '05: Proceedings of the 2005 ACM/IEEE Conference on Supercomputing, IEEE, (2005).

\bibitem{Schollwock2005} 
U. Schollw\"{o}ck,
The density-matrix renormalization group
Rev. Mod. Phys. \textbf{77}, 259 (2005).

\bibitem{Stoudenmire2011} 
E. M. Stoudenmire and S. R. White,
Studying two dimensional systems with the density matrix renormalization group,
arXiv:1105.1374.


\bibitem{Somma2002}
R. Somma, G. Ortiz, J. E. Gubernatis, E. Knill, and R. Laflamme,
Simulating physical phenomena by quantum networks,
Phys. Rev. A \textbf{65}, 042323 (2002).

\bibitem{Wecker2015s} 
D. Wecker, M. B. Hastings, N. Wiebe, B. K. Clark, C. Nayak, and M. Troyer,
Solving strongly correlated electron models on a quantum computer,
Phys. Rev. A \textbf{92}, 062318 (2015).

\bibitem{Wecker2015p} 
D. Wecker, M. B. Hastings, and M. Troyer, 
Progress towards practical quantum variational algorithms,
Phys. Rev. A \textbf{92}, 042303 (2015).

\bibitem{Jiang2018}
Z. Jiang, K. J. Sung, K. Kechedzhi, V. N. Smelyanskiy, and S. Boixo, 
Quantum algorithms to simulate many-body physics of correlated fermions,
Phys. Rev. Appl. \textbf{9}, 044036 (2018).

\bibitem{Reiner2019} 
J.-M. Reiner, F. Wilhelm-Mauch, G. Sch\"{o}n, and M. Marthaler, 
Finding the ground state of the Hubbard model by variational methods on a quantum computer with gate errors,
Quantum Sci. Technol. \textbf{4}, 035005 (2019).

\bibitem{Cade2020} 
C. Cade, L. Mineh, A. Montanaro, and S. Stanisic,
Strategies for solving the Fermi-Hubbard model on near-term quantum computers,
Phys. Rev. B \textbf{102}, 235122 (2020).

\bibitem{Cai2020} 
Z. Cai, 
Resource estimation for quantum variational simulations of the Hubbard model,
Phys. Rev. Appl. \textbf{14}, 014059 (2020).

\bibitem{Demers2020} 
P.-L. Dallaire-Demers, M. St\k{e}ch\l{}y, J. F. Gonthier, N. T. Bashige, J. Romero, and Y. Cao, 
An application benchmark for fermionic quantum simulations,
arXiv:2003.01862.

\bibitem{Martin2022} 
B. A. Martin, P. Simon, and M. J. Ran\v{c}i\'{c},
Simulating strongly interacting Hubbard chains with the variational Hamiltonian ansatz on a quantum computer,
Phys. Rev. Research \textbf{4}, 023190 (2022). 

\bibitem{Gard2022} 
B. T. Gard and A. M. Meier,
A classically efficient quantum scalable Fermi-Hubbard benchmark,
Phys. Rev. A \textbf{105}, 042602 (2022). 

\bibitem{Stanisic2022}
S. Stanisic, J. L. Bosse, F. M. Gambetta, R. A. Santos, W. Mruczkiewicz, T. E. O'Brien, E. Ostby,
and A. Montanaro,
Observing ground-state properties of the Fermi-Hubbard model using a scalable algorithm on 
a quantum computer,
Nat. Commun. \textbf{13}, 5743 (2022).

\bibitem{Ayral2023} 
T. Ayral, P. Besserve, D. Lacroix, and E. A. R. Guzman, 
Quantum computing with and for many-body physics,
Eur. Phys. J. A \textbf{59}, 227 (2023).

\bibitem{Cerezo2021} 
M. Cerezo, A. Arrasmith, R. Babbush, S. C. Benjamin, S. Endo, K. Fujii, J. R. McClean, K. Mitarai,
X. Yuan, L. Cincio, P. J. Coles,
Variational quantum algorithms,
Nat. Rev. Phys. \textbf{3}, 625 (2021).

\bibitem{Alexeev2024}
Y. Alexeev, M. Amsler, M. A. Barroca, \textit{et al}., 
Quantum-centric supercomputing for materials science: A perspective on challenges and future directions, 
Future Generation Computer Systems \textbf{160}, 666 (2024).

\bibitem{Farhi2014} 
E. Farhi, J. Goldstone, and S. Gutmann,
A quantum approximate optimization algorithm, 
arXiv:1411.4028.

\bibitem{Tilly2022} 
J. Tilly, H. Chen, S. Cao, D. Picozzi, K. Setia, Y. Li, E. Grant, L. Wossnig, I. Rungger, 
G. H. Booth, J. Tennyson,
The Variational Quantum Eigensolver: A review of methods and best practices,
Phys. Rep. \textbf{986}, 1 (2022).

\bibitem{Blekos2024} 
K. Blekos, D. Brand, A. Ceschini, C.-H. Chou, R.-H. Li, K. Pandya, A. Summer,
A review on Quantum Approximate Optimization Algorithm and its variants,
Phys. Rep. \textbf{1068}, 1 (2024).

\bibitem{Nielsen2010}
M. A. Nielsen and I. L. Chuang, 
\textit{Quantum Computation and Quantum Information},
(Cambridge University Press, Cambridge, UK, 2010).

\bibitem{Kitaev1995}
A. Yu. Kitaev, 
Quantum measurements and the Abelian Stabilizer Problem,
arXiv:quant-ph/9511026.

\bibitem{Abrams1997} 
D. S. Abrams and S. Lloyd, 
Simulation of many-body fermi systems on a universal quantum computer,
Phys. Rev. Lett. \textbf{79}, 2586 (1997).

\bibitem{Abrams1999} 
D. S. Abrams and S. Lloyd, 
Quantum algorithm providing exponential speed increase for finding eigenvalues and eigenvectors,
Phys. Rev. Lett. \textbf{83}, 5162 (1999).

\bibitem{Guzik2005} 
A. Aspuru-Guzik, A. D. Dutoi, P. J. Love, and M. Head-Gordon, 
Simulated quantum computation of molecular energies,
Science \textbf{309}, 1704 (2005).

\bibitem{Whitfield2011} 
J. D. Whitfield, J. Biamonte, and A. Aspuru-Guzik, 
Simulation of electronic structure Hamiltonians using quantum computers,
Mol. Phys. \textbf{109}, 735 (2011).

\bibitem{Parker2000} 
S. Parker and M. B. Plenio,
Efficient factorization with a single pure qubit and $\log N$ mixed qubits,
Phys. Rev. Lett. \textbf{85}, 3049 (2000).

\bibitem{Dobsicek2007} 
M. Dob\v{s}\'{\i}\v{c}ek, G. Johansson, V. Shumeiko, and G. Wendin,
Arbitrary accuracy iterative quantum phase estimation algorithm using a single ancillary qubit: A two-qubit benchmark,
Phys. Rev. A \textbf{76}, 030306(R) (2007).

\bibitem{Mei2007} 
L. Xiu-Mei, L. Jun, and S. Xian-Ping, 
Experimental realization of arbitrary accuracy iterative phase estimation algorithms on ensemble quantum computers,
Chinese Phys. Lett. \textbf{24}, 3316 (2007).

\bibitem{Qiskit}
Qiskit, https://quantum.cloud.ibm.com/docs/en/guides

\bibitem{Qis-Aer}
A. Javadi-Abhari, M. Treinish, K. Krsulich, C. J. Wood, J. Lishman, J. Gacon, S. Martiel, 
P. D. Nation, L. S. Bishop, A. W. Cross, B. R. Johnson, and J. M. Gambetta, 
Quantum computing with Qiskit, arXiv:2405.08810.

\bibitem{IBM}
IBM Quantum Platform, https://quantum.cloud.ibm.com/



\bibitem{Geim2007}
A. K. Geim and K. S. Novoselov,
The rise of graphene ,
Nat. Mater. \textbf{6}, 183 (2007).

\bibitem{Castro2009} 
A. H. Castro Neto, F. Guinea, N. M. R. Peres, K. S. Novoselov, and A. K. Geim, 
The electronic properties of graphene,
Rev. Mod. Phys. \textbf{81}, 109 (2009).

\bibitem{Fernandez2007} 
J. Fern\'andez-Rossier and J. J. Palacios,
Magnetism in graphene nanoislands,
Phys. Rev. Lett. \textbf{99}, 177204 (2007).

\bibitem{Rozhkov2011} 
A. V. Rozhkov, G. Giavaras, Y. P. Bliokh, V. Freilikher, and F. Nori,
Electronic properties of mesoscopic graphene structures: Charge confinement and control of spin and charge transport,
Phys. Rep. \textbf{503}, 77 (2011).

\bibitem{Yazyev2010}
O. V. Yazyev, 
Emergence of magnetism in graphene materials and nanostructures,
Rep. Prog. Phys. \textbf{73}, 056501 (2010).

\bibitem{Guclu2014} 
A. D. G\"{u}\c{c}l\"{u}, P. Potasz, M. Korkusinski, and P. Hawrylak,
Graphene quantum dots,
(Springer, Berlin Heidelberg 2014).

\bibitem{Oteyza2022}
D. G. de Oteyza and T. Frederiksen,
Carbon-based nanostructures as a versatile platform for tunable $\pi$-magnetism,
J. Phys.: Condens. Matter \textbf{34}, 443001 (2022).

\bibitem{Ortiz2001}
G. Ortiz, J. E. Gubernatis, E. Knill, and R. Laflamme,
Quantum algorithms for fermionic simulations,
Phys. Rev. A \textbf{64}, 022319 (2001).


\bibitem{Kivlichan2018}
I. D. Kivlichan, J. McClean, N. Wiebe, C. Gidney, A. Aspuru-Guzik, G. K. -L. Chan, and R. Babbush, 
Quantum simulation of electronic structure with linear depth and connectivity,
Phys. Rev. Lett. \textbf{120}, 110501 (2018).

\bibitem{Farhi2001}
E. Farhi, J. Goldstone, S. Gutmann, J. Lapan, A. Lundgren, and D. Preda, 
A quantum adiabatic evolution algorithm applied to random instances of an NP-complete problem,
Science \textbf{292}, 472 (2001).

\bibitem{QisNat} 
Qiskit Nature, https://doi.org/10.5281/zenodo.7828767v

\bibitem{Hatano2005} %
N. Hatano and Masuo Suzuki,
``Finding exponential product formulas of higher orders" 
In \textit{Quantum annealing and other optimization methods}, 
(Springer, Berlin, Heidelberg, 2005) pp. 37-68.

\bibitem{Jordan1928}
P. Jordan and E. Wigner, 
\"{U}ber das Paulische \"{A}quivalenzverbot,
Z. Phys. A. \textbf{47}, 631 (1928).

\bibitem{QuSpin}
P. Weinberg and M. Bukov,
QuSpin: a Python package for dynamics and exact diagonalisation of quantum many body systems. 
Part II: bosons, fermions and higher spins,
SciPost Phys. \textbf{7}, 020 (2019)·

\end{thebibliography}
\end{document}